\documentclass[%
twocolumn,
superscriptaddress,
amsmath,amssymb
]{revtex4-2}

\usepackage{graphicx,color,hyperref,xcolor}
\usepackage{dcolumn}
\usepackage{bm}
\usepackage{hyperref}

\newcommand{\bra}[1]{\langle #1|}
\newcommand{\ket}[1]{|#1\rangle}
\newcommand{\braket}[2]{\langle #1|#2\rangle}

\begin{document}
	\title{Quantum simulation of parity-time symmetry breaking with a superconducting quantum processor}

	\author{Shruti Dogra}
	\email{shruti.dogra@aalto.fi}
	\affiliation{QTF Centre of Excellence, Department of Applied Physics, Aalto University School of Science, P.O. Box 15100, FI-00076 AALTO, Finland.}

         \author{Artem A. Melnikov}\altaffiliation{Also at: Terra Quantum AG, St. Gallerstrasse 16A, 9400 Rorschach, Switzerland.}
         \affiliation{QTF Centre of Excellence, Department of Applied Physics, Aalto University School of Science, P.O. Box 15100, FI-00076 AALTO, Finland.}
     \affiliation{Physics Department, Moscow Institute of Physics and Technology,  Institutskii  per.  9,  Dolgoprudny,  141700,  Moscow  District,  Russia.}
	\author{Gheorghe Sorin Paraoanu} 
	\email{sorin.paraoanu@aalto.fi}
	\affiliation{QTF Centre of Excellence, Department of Applied Physics, Aalto University School of Science, P.O. Box 15100, FI-00076 AALTO, Finland.}

	\date{\today}
	
\begin{abstract}
\textbf{Abstract- }
The observation of genuine quantum effects in systems governed by non-Hermitian Hamiltonians has been an outstanding challenge in the field. Here we simulate the evolution under such Hamiltonians in the quantum regime on a superconducting quantum processor by using a dilation procedure involving an ancillary qubit. We observe the parity-time ($\mathcal{PT}$)-symmetry breaking phase transition at the exceptional points, obtain the critical exponent, and show that this transition is associated with a loss of state distinguishability. In a two-qubit setting, we show that the entanglement can be modified by local operations. 	
\end{abstract}
     	\maketitle

\section*{Introduction  \label{sec:intro}}
The Hermiticity of physical observables is a fundamental tenant of standard quantum physics,
guaranteeing real eigenspectra and leading to the generation of unitary dynamics in closed quantum systems. 
However, this is needlessly restrictive: it has been shown \cite{Bender1998} that non-Hermitian Hamiltonians 
endowed with parity and time ($\mathcal{PT}$) symmetry possess real positive eigenvalues and eigenvectors with positive norm. 
Experimental platforms where non-Hermitian Hamiltonians can be implemented comprise optical waveguides ~\cite{christian-nature-2010,Szameit2019,Klauck2019}, polarized photons ~\cite{xiao-nature-2017}, nuclear spins \cite{Lu2013, wen-pra-2019}, superconducting circuits \cite{Zueco2018,Murch2019}, mechanical oscillators ~\cite{Bender2013a}, nitrogen-vacancy centers in diamond \cite{pick-prr-2019,yang-science-2019}, fiber-optics networks \cite{Regensburger2012}, and ultracold Fermi gases \cite{Luo2019}. Open systems are a natural candidate for realizing these Hamiltonians, since non-Hermitian terms appear naturally as a consequence of energy being injected or lost \cite{ramy-nature-2018}. 
 
However, a  major drawback of open-system approaches 
is the need to control precisely the gain and the dissipation: these experiments require complicated setups with gain and alternating losses  \cite{xiao-nature-2017}, and yet only wave-like effects can be observed.
Moreover, employing gain-loss systems for the study of quantum properties such as entropy, entanglement, and correlations is fundamentally impossible because gain inevitably adds noise \cite{Scheel_2018}. Thus, in order to make progress one would need genuine realizations of non-Hermitian dynamics in the quantum regime \cite{Murch2019, yang-science-2019,Klauck2019}, which  maintain and allow the measurement of delicate quantum effects.

Here we show that the non-Hermitian dynamics can be simulated digitally \cite{Paraoanu2014} in a superconducting quantum processor by extending the 
Hilbert space with the use of an ancilla qubit and under the action of appropriately-defined gates. 
To achieve this, 
we combine two techniques: a dilation method which is universal (applicable to any Hamiltonian) \cite{yang-science-2019} 
and an optimal method for generating any two-qubit gate with combinations of single qubit gates and at most three CNOT 
gates \cite{vatan-pra-2004,yang-npjqi-2019}.  
This combination enables us to observe and fully characterize the
broken $\mathcal{PT}$-symmetry transition and to settle decisively the relationship between non-Hermitian quantum mechanics and no-go theorems on state distinguishability and monotony of entanglement \cite{Bender2013,chen-pra-2014,kohei-prl-2017,Nori2019}. 
We achieve this by making use of the emergent technology of superconducting processors, on which significant technical progress has been shown in recent times by IBM \cite{Gambetta2019}. Although still imperfect, these devices have already enabled important results such as quantum error correction \cite{PhysRevA.94.032329}, fault-tolerant gates \cite{PhysRevLett.122.080504}, proofs of violation of Mermin \cite{PhysRevA.94.012314} and Leggett-Garg \cite{PhysRevA.95.032131} inequalities, demonstrations of non-local parity measurements \cite{Panigrahi2019,Paraoanu2018},
simulations of paradigmatic models in open quantum systems \cite{Maniscalco2020}, the creation of highly entangled graph states \cite{Zeng2018}, the determination of molecular ground-state energies \cite{Gambetta2017}, the implementation of quantum witnesses \cite{Nori_cats_2019}, and quantum-enhanced solutions to large systems of linear equations  \cite{Perelshtein2020}. The use of a superconducting quantum processor offers the possibility of extracting all relevant quantum correlations and of designing and programming efficiently the required gates, adapted to the particular topology of the machine.

\par

We consider the generic system qubit non-Hermitian Hamiltonian \cite{Bender2007} with natural units $\hbar =1$
\begin{equation}
H_{\mathrm q} = \sigma_{x} + ir \sigma_{z}, \label{eq:hamiltonian}
\end{equation}
where $r$ is a real parameter and $\sigma_{x}$ and $\sigma_{z}$ are the Pauli matrices. The eigenvalues 
are $\pm \sqrt{1-r^2}$, and the condition for non-Hermiticity is simply $r\neq 0$. The parity 
operator is $\mathcal{P} = \sigma_{x}$ and the time-inversion operator is the complex conjugation operator 
$\mathcal{T} = \star$. This Hamiltonian has an exceptional 
point at $|r| = 1$, where the two eigenvectors coalesce and the eigenvalues become parallel. For $|r| < 1$ the 
eigenvalues are real, corresponding to distinct eigenvectors, and the Hamiltonian satisfies $\mathcal{PT}$-symmetry
$[\mathcal{PT}, H_{\mathrm q}] = 0$ (see Supplementary Note \ref{sm:hamiltonian});
for $|r| > 1$, the eigenvalues become imaginary and the $\mathcal{PT}$-symmetry is broken. The Hamiltonian 
Eq.~(\ref{eq:hamiltonian}) can be understood as the standard non-Hermitian form providing equal-coupling 
(off-diagonal terms) between the basis states as well as equal gain and loss via the complex diagonal terms 
required for $\mathcal{PT}$-symmetry \cite{Alu2019}.

We realize the Hamiltonian Eq. (\ref{eq:hamiltonian}) in a dilated space 
	with the help of an ancilla, observing single-qubit dynamics under PT-symmetric Hamiltonians and witnessing the coalesce of eigenvectors at the exceptional points. Further, by
allowing different quantum states to evolve under the same set of operations generated by non-
Hermitian generators, we show that the trace distance between arbitrary states is modified - a task
that is forbidden in Hermitian quantum mechanics. We extract the critical exponent of the transition, obtaining a value in agreement with theoretical predictions. We also observe an apparent violation of entanglement monotonicity in a two-qubit system, where one of the qubits is driven by a non-Hermitian Hamiltonian. Finally, we conclude by 
providing the complete dynamics of correlations developed between system qubits and the ancilla.

\begin{widetext}

\begin{figure}
\includegraphics[scale=1]{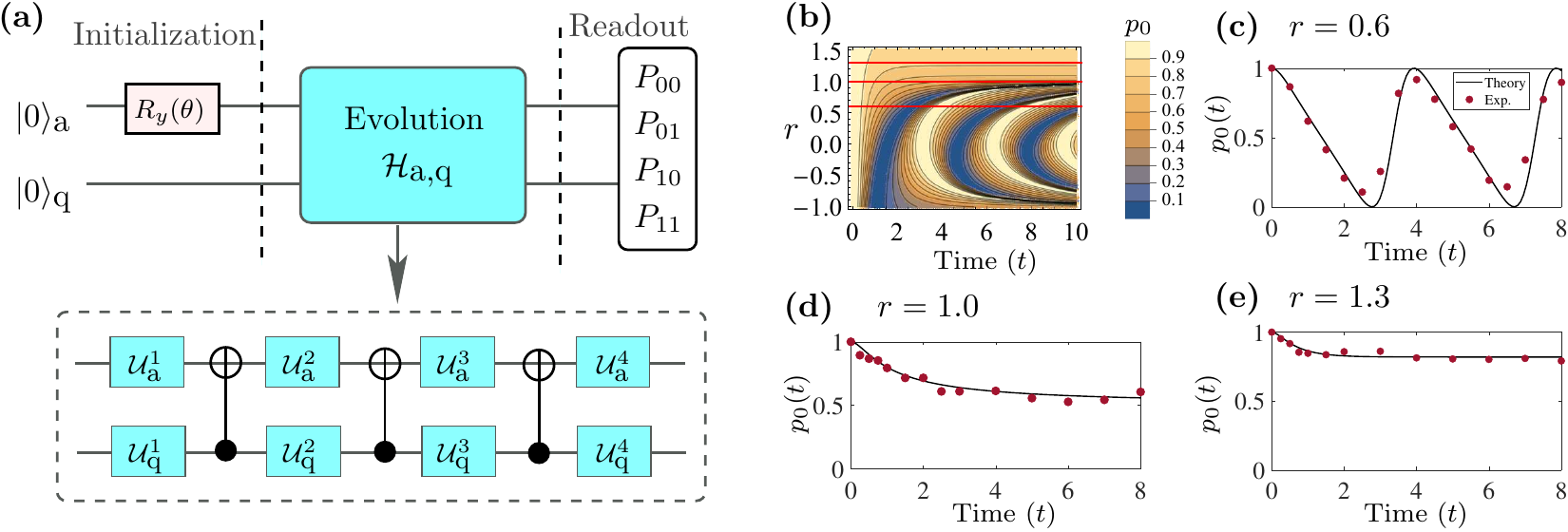}
\caption{{\bf Demonstration of $\mathcal{PT}$-symmetry breaking for a single-qubit.}
(a) Quantum circuit implementing the dilated unitary operator $U_{\mathrm a,q}$ in the four-dimensional Hilbert space
of system and ancilla, where $R_y(\theta)$ is a single-qubit rotation about the $y$ axis by an angle $\theta$, and the 
subscript $\mathrm q$($\mathrm a$) refers to the system qubit(ancilla). Details of the evolution under $\mathcal{H}_{\mathrm{a,q}}$ 
are given in the inset. (b) Dynamics of the population $p_{0}$ of state $|0\rangle$ under the Hamiltonian $H_{\mathrm q}$ for a 
range of parameters $-1 \leq r \leq 1.5$. As the eigenvectors of $H_{\mathrm q}$ coalesce at $r=1$, the oscillations disappear, 
and beyond this point, $\mathcal{PT}$-symmetry breaking is observed. The plots (c),(d),(e) present the results of the 
experiments (red dots) run on the IBM quantum experience (with $8192$ repetitions) for the parameter $r$ taking values $r=0.6$, $1.0$, and $1.3$ 
respectively. The data matches well with theory (continuous black lines) corresponding to the three red lines from (b). 
} \label{fig:pop}
\end{figure}

\end{widetext}
\section*{Results   \label{sec:results}}

To realize the Hamiltonian Eq. (\ref{eq:hamiltonian}) we use a Naimark dilation procedure employing an additional ancilla qubit and a Hermitian operator $\mathcal{H}_{\mathrm a,q}(t)$ acting on the total qubit-ancilla Hilbert space. The dynamics under 
$\mathcal{H}_{\mathrm a,q}(t)$ is determined by the Schr\"odinger equation
\begin{equation}
i\frac{d}{dt}\ket{\Psi (t)}_{\mathrm a,q} = \mathcal{H}_{\mathrm a, q}(t)\ket{\Psi(t)}_{\mathrm a,q}, \label{eq:extendedSchr}
\end{equation}
whose solution is given by
\begin{equation}\label{bigpsi}
\ket{\Psi(t)}_{\mathrm a, q} = \ket{0}_{\mathrm a}\ket{\psi(t)}_{\mathrm q} + \ket{1}_{\mathrm a}\ket{\tilde{\psi}(t)}_{\mathrm q}, 
\end{equation}
 where $\ket{\psi(t)}_{\mathrm q}$ is the solution of  $i\frac{d}{dt}\ket{\psi(t)}_{\mathrm q}= H_{\mathrm q}\ket{\psi(t)}_{\mathrm q}$. 
Here $\ket{\tilde{\psi}(t)}_{\mathrm q}=\eta(t)\ket{\psi(t)}_{\mathrm q}$, where $\eta(t)$ is a positive linear operator given by
$ \eta(t)=\left[ (1+\eta_0^2) \exp(-iH_{\mathrm q}^{\dag}t)\exp (i H_{\rm q}t) -\mathbb{I} \right]^{1/2}$~\cite{yang-science-2019},
with $\eta(0)=\eta_0 \mathbb{I}$ at the initial time $t=0$ (see Methods).

In order to obtain a solution of the form Eq.~(\ref{bigpsi}), the ancilla and the qubit 
are initialized in the state $\ket{\Psi(0)}_{\mathrm a, q} = (\ket{0}_{\mathrm a} + \eta_0 \ket{1}_{\mathrm a}) \otimes \ket{\psi(0)}_{\mathrm q}$,
as shown in Fig.~\ref{fig:pop}(a), by using a rotation $R_{y}(\theta )$ 
on the ancilla qubit,
where $\theta=2\tan^{-1}\eta_0$ and $\ket{\psi(0)}_{\mathrm q}=\ket{0}_{\mathrm q}$. Thus, the dynamics of the system qubit in the subspace with the ancilla in state $\ket{0}_{\mathrm a}$ satisfies 
the desired evolution by the non-Hermitian Hamiltonian Eq. (\ref{eq:hamiltonian}). For an arbitrary time $t$ and 
for any given $r$ the corresponding unitary operator $U_{\mathrm a,q}(t)= \mathrm {T} \exp [-i\int_{0}^{t} \mathcal{H}_{\mathrm a,q}(\tau)d\tau]$ 
is obtained numerically as described below.

\subsection*{Parity-time symmetry breaking in a single-qubit}

We implement the unitary operator $U_{\mathrm a,q}(t)$ using the q[0] and q[1] qubits of a five-qubit
IBM quantum processor for different values of the Hamiltonian parameter $r$. 
We start by presenting in Fig.~\ref{fig:pop}(b) the expected theoretical results for the ground-state population obtained under $H_{\mathrm{q}}$, 
\textit{i.e.} generated by the non-unitary evolution operator $\exp (-i H_{\mathrm{q}} t)$ (see Methods). The breaking of  $\mathcal{PT}$-symmetry, as one crosses the 
exceptional point $r = 1$, is clearly visible. Indeed, for $r < 1$ the eigenvalues of $H_{\mathrm q}$ are real and one observes Rabi oscillations. When  
$|r|$ exceeds $1$, the eigenvalues of $H_{\mathrm q}$ become imaginary, the $\mathcal{PT}$-symmetry is broken, and what 
one observes is the decay of the population. Fig.~\ref{fig:pop}(c-e) present the results from the experimental realization
of $H_{\mathrm q}$ on IBM quantum experience for three different values of $r$. We note that the agreement with the theoretical values is excellent. Each experiment is repeated $8192$ times.  Thus, the statistical errors here are of the order of $1/\sqrt{8192} = 0.01$, which lead to the error bars too small to be shown distinctly in the experimental plots. 	In addition, in various experiments presented here,
	we track the systematic errors in terms of 
	measurement corrections and incorporate these corrections in  respective experimental datasets as described in Methods, with further details given in the Supplementary Note \ref{supp-note-5}.

The details of the implementation are shown in Fig.~\ref{fig:pop}(a). We start with the qubit and the ancilla both
initialized in the state $\vert 0 \rangle$, after which the ancilla alone is subjected to a rotation along the $y$-axis by an angle $\theta$, which initializes the ancilla subspace $\theta$~\cite{yang-science-2019}.
The explicit form of the operator $U_{\mathrm a,q}(t)$ at any arbitrary time $t$ is found by a numerical decomposition into 
single and two-qubit gates~\cite{vatan-pra-2004}. This decomposition $U_{\mathrm{num}}(t)=U_{n} \dots U_{1}$
matches with the desired unitary operator $U_{\mathrm a,q}(t)$ 
with a fidelity $F_U>0.99$, where the fidelity is defined as $F_U=1-||U_{\mathrm{num}}(t)-U_{\mathrm a,q}(t)||/||U_{\mathrm a,q}(t)||$ (also see Supplementary Note \ref{sm:operator}).
The quantum circuit which implements the decomposition of $U_{\mathrm{num}}(t)$ (see inset of Fig.~\ref{fig:pop}(a)) comprises  a sequence of single qubit rotations $\mathcal{U}_{\mathrm q(a)}^j$, each of them having up to three degrees 
of freedom, and three two-qubit controlled-NOT gates~\cite{vatan-pra-2004, yang-npjqi-2019}. The width of this circuit is 2 and the depth is 8. Specifically, the single-qubit gates are general rotations given by $\mathcal{U}_{\mathrm q(a)}^j(\alpha, \beta, \gamma)=R_z(\alpha)_{\mathrm q(a)}^j R_y(\beta)_{\mathrm q(a)}^jR_z(\gamma)_{\mathrm q(a)}^j$,
where $\alpha$, $\beta$, $\gamma$ are the angles of rotations and the operators $R_y$, $R_z$ correspond to the rotations generated by the Pauli operators $\sigma_y$ and $\sigma_z$ respectively. The operator $\mathcal{U}_{\mathrm q(a)}^j(\alpha,\beta,\gamma)$ has a direct correspondence with the single-qubit operator U3, as defined by IBM.
For instance, given $r=0.6$ and $t=0.5$ (see Fig.~\ref{fig:pop}(a)), we have the following set of operations: 
$\mathcal{U}_\mathrm{a}^1(2.83, 0.55, 3.72), \, \mathcal{U}_{\mathrm q}^1(0.51, -2.98, 1.63)$, 
$\mathcal{U}_\mathrm{a}^2(-1.75, -3.34, -4.60), \, \mathcal{U}_{\mathrm q}^2(0.00, 0.00, 4.02)$, 
$\mathcal{U}_\mathrm{a}^3(4.81, 3.08, -1.02), \, \mathcal{U}_{\mathrm q}^3(0.01, 0.29, 0.04)$, and
$\mathcal{U}_\mathrm{a}^4(0.00, -5.19, 0.50), \, \mathcal{U}_{\mathrm q}^4(0.46, -1.51, 0.37)$.
After the $U_{\mathrm{num}}(t)$ implementation, the post-selected 
subspace of our interest corresponds to the ancilla in state $\vert 0 \rangle_a$.
At the end of the algorithm we measure the probabilities $P_{kl}$ of the qubit-ancilla state in the computational basis $\{ |kl\rangle_{\mathrm{a,q}} \equiv |k\rangle_{\mathrm{a}} |l\rangle_{\mathrm{q}} \}$ with $k,l\in\{ 0,1\}$. Finally, the 
ground state population in the desired post-selected subspace of the system-qubit is given by, $p_0(t)=P_{00}/(P_{00}+P_{01})$,
which can be obtained directly from the experiments. These are shown with red dots in Fig.~\ref{fig:pop}(c-e) and follow very closely the results for the population in the $|0\rangle_{\mathrm{q}} $ state of the qubit under the non-Hermitian Hamiltonian Eq. (\ref{eq:hamiltonian}). The results demonstrate a high-fidelity simulation of the $\mathcal{PT}$-symmetry breaking in a single qubit.
\subsection*{Quantum state distinguishability}
\begin{figure}
 \centering
 \includegraphics[scale=1]{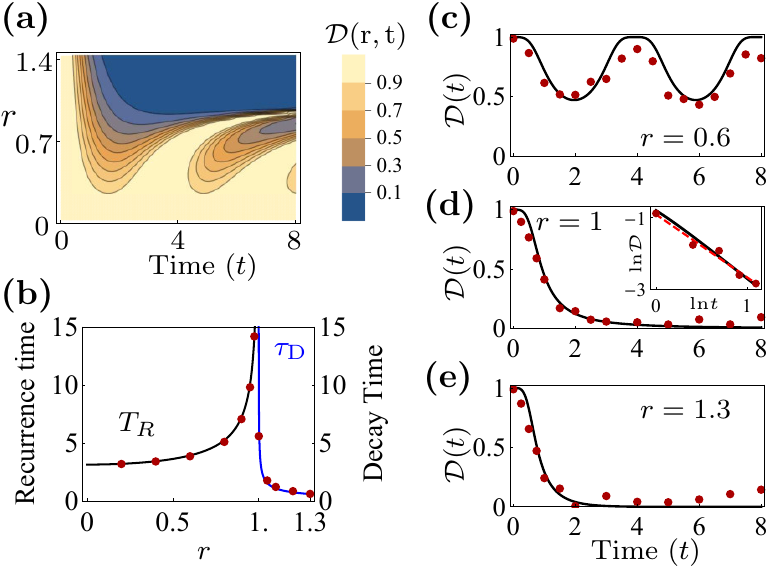}
 \caption{{\bf Quantum state distinguishability in a system driven by  a non-Hermitian Hamiltonian.}
 (a) Variation of trace distance Eq.~(\ref{eq:dist}) for single-qubit initial states $\rho_{1\rm q}(t=0)=|0\rangle \langle0|$ and $\rho_{2\rm q}(t=0)=|1\rangle \langle1|$,
 as they evolve under $H_{\rm q}$. (b) Experimental (red dots) and theoretical recurrence (continuous black line) and decay times
 (continuous blue line) characterizing retrieval and loss of information in the two phases. The plots in (c)-(e) present the theoretical (continuous black line) and experimental (red dots) curves  
 of $\mathcal{D}(\rho_{1\rm q}(\rm{t}), \rho_{2\rm q}(\rm{t}))$ for $\rm{r}=0.6$, $1.0$ and $1.3$ respectively. The inset in (d) shows $\ln \mathcal{D}$ versus $\ln t$ for $t\in [1,3]$, where the dotted red line is a linear interpolation of the data.
 Each experimental point is obtained by averaging over $8192$ samples,
 such that the errors are below the size of the marker.
  \label{fig:dist1}}
\end{figure}

Next, we demonstrate an unexpected consequence of non-Hermitian dynamics concerning state distinguishability.
Designing a general protocol to distinguish two (or more) arbitrary quantum states is a challenge in standard Hermitian quantum mechanics. On the other hand, the evolution of an arbitrary pair of states under a non-Hermitian operator can alter the distance between them, and may even make the arbitrary pair of quantum states orthogonal~\cite{Bender2013,kohei-prl-2017,Nori2019}. To observe this unusual feature of non-Hermitian dynamics we use the quantum circuit in Fig.~\ref{fig:pop}(a) to evolve the system qubit, initialized respectively in the orthogonal states $\vert 0 \rangle_{\rm q}$ and $\vert 1 \rangle_{\rm q}$. At various different instances of time, the state of the system qubit in the postselected sub-space with ancilla in state $\vert 0 \rangle_{\rm a}$ is obtained and the trace distance 
\begin{equation} 
\mathcal{D}(\rho_{1\rm q}(\rm{t}), \rho_{2\rm q}(\rm{t})) = \frac{1}{2} \rm{tr} \sqrt{\rho_{\rm diff}(\rm{t})^{\dag} \, \rho_{\rm diff}(\rm{t})}, \label{eq:dist}
\end{equation}
between the respective states is determined, where $\rho_{\rm diff}(\rm{t}) = \rho_{1\rm q}(\rm{t}) - \rho_{2\rm q}(\rm{t})$ and $\rho_{i\rm q}(t)=|\psi_i(t)\rangle_{\rm q} \langle \psi_i(t)|_{\rm q}$. For the given pair of initial states, the expected pattern for the variation of $\mathcal{D}$ with r and t is shown in
Fig.~\ref{fig:dist1}(a). The characteristic recurrence time $T_R$ in the $\mathcal{PT}$-symmetric phase and the decay time $\tau_{\rm D}$ in the broken-symmetry phase are plotted in Fig.~\ref{fig:dist1}(b) and compared to their analytical expressions ($T_R=\pi/\sqrt{1-r^2}$ and $\tau_{\rm D}=1/2\sqrt{r^2-1}$ from Supplementary Equations (\ref{eq:ReccTime}) and (\ref{eq:DecayTime})). Note that these times reflect the delicate balance between gain and loss, which is encoded in the structure of the Hamiltonian (see Supplementary Note \ref{supp-note-4}). 
In Fig.~\ref{fig:dist1}(c)-(e) we show the experimentally-obtained variation of the trace distance for three different values of r. An oscillating pattern in the 
trace distance is obtained for r$<1$, which is a signature of information exchange between the system and the environment, while for r$\geq 1$ we measure a decay pattern, which corresponds to 
loss of information to the environment.
Interestingly, the oscillations in distinguishability correspond to oscillations in entanglement of qubit-ancilla state ~\cite{kohei-prl-2017} (see Supplementary Fig. \ref{fig:norm}). 
For $r=1$ (exceptional point) these timescales diverge, and one cannot define anymore a characteristic time of the system. Instead, in close analogy with phase transitions in many-body systems, the distinguishability follows asymptotically a power-law $\mathcal{D} \sim t^{-\delta}$ \cite{kohei-prl-2017}, where the critical exponent $\delta = 2$ corresponds to two coalescing eigenstates.  We have first checked numerically that for $t \gg 1$ the distinguishability indeed displays this power-law behavior, with the critical exponent very close to 2. We can verify this scaling also experimentally, with the caveat that for $t\gtrsim 3$ the distingusihability becomes already smaller than the precision that we can reach on the IBM machine. Still, we can identify an interval $t \in [1,3]$ where the theoretical plot $\ln \mathcal{D}$ versus $\ln t$ starts to be approximately linear, 
with slope $\delta = 1.93 \pm 0.08$, see inset of Fig.~\ref{fig:dist1}(d). In this region, we obtain by fitting the experimental data $\delta = 1.75 \pm 0.15$ (dashed red line in the inset), a reasonably close value.

Evaluating the distinguishability requires a 
	complete characterization of the single-qubit density matrices 
	$\rho_{1\rm{q}}(t)$ and $\rho_{2\rm{q}}(t)$, which is done by
	a set of three operations that independently fetch the elements of the density matrix. Each of these experiments is
	repeated $8192$ times, such that even after evaluating Eq.~\ref{eq:dist}, the  statistical error in the measure
	of distinguishability remains small.

\subsection*{Bipartite systems under non-Hermitian evolution}
Next, we observe the dynamics of entanglement in a bipartite-system when one of the parties undergo a
local operation generated by $H_{\mathrm q}$, for different values of $r$. Such scenarios have been studied theoretically
~\cite{chen-pra-2014,kohei-prl-2017},
and it was shown that entanglement restoration and information recovery can happen in the $\mathcal{PT}$-symmetric phase. This breaks entanglement monotonicity, allowing the creation of entanglement by a local operation. This unusual effect is due to the modified evolution in the postselected subspace due to mere existence of a component of the total wavefunction outside this subspace  \cite{Klauck2019,paraoanu2011,paraoanu2011foundations}. 

To study this phenomenon, we consider a system consisting of two qubits $\rm q$ and $\rm q'$ initialized in a maximally entangled Bell state, 
$\vert \Phi^{+} \rangle_{\rm q, q'}=(\vert 00 \rangle_{\rm q, q'} + \vert 11 \rangle_{\rm q, q'})/\sqrt{2}$. 
One system qubit (say $\rm q$) undergoes a non-Hermitian evolution by $\mathbb{H}_{\rm q,q'}=H_{\mathrm q} \otimes \mathbb{I}_{\mathrm q'}$  
with the help of an ancillary qubit $\mathrm{a}$, such that the total Hamiltonian including the dilation is  
$\mathbb{H}_{\rm a,q,q'}=\mathcal{H}_{\rm a,q} \otimes \mathbb{I}_{\mathrm q'}$ leading to a unitary evolution, $\mathbb{U}_{\rm a,q,q'}=U_{\rm a,q} \otimes \mathbb{I}_{\mathrm q'}$. Finally, the three-partite state of the system is measured and postselected subject to the state of the ancilla being $\vert 0 \rangle_{\mathrm a}$.

\begin{widetext}

\begin{figure}[h!]
\centering
\includegraphics[scale=1]{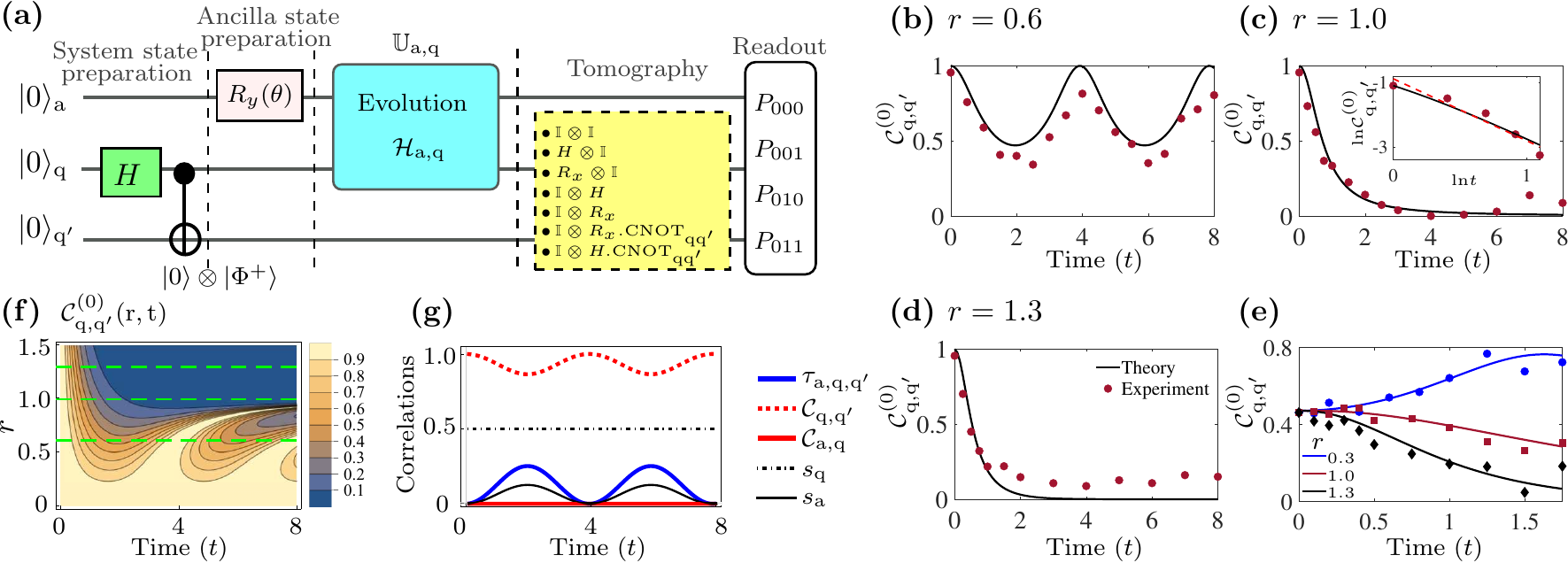}
\caption{{\bf Demonstration of an apparent violation of entanglement monotonicity.}
(a) Quantum circuit implemented on the superconducting quantum processor, where the qubit ${\rm a}$ serves as ancilla and ${\rm q}$ and ${\rm q'}$ form a qubit-qubit bipartite system.
In (b)-(d) we present the results from the experiments, where the variation of concurrence with time is shown for different values of $r$, with each experiment being repeated $8192$ times. 
Experimental data (red dots) very closely
follow the theoretically expected behaviour (continuous black lines). The inset in (c) presents the variation of the logarithm of concurrence versus 
$\ln t$ in the interval $t\in [1,3]$, where the dashed red line is a linear fit to the experimental red circles. (e) We prepare the qubits q and q' in a state with concurrence 0.475 and we evolve the system under the local non-Hermitian evolution with up to $t=1.75$, for  different values of $r$. Solid blue, red and black curves present the theoretically expected results while blue circle, red square and black diamond
markers correspond to the respective experimental measurements.
In (f) we show the theoretical plots for the variation of concurrence with time, with one of the qubits undergoing a local non-Hermitian evolution, for various parameters $r \in [0, 1.5]$; the horizontal green lines mark
the values of $r$ corresponding to the experimental results in (b)-(d).
The correlations among the system qubits and ancilla, simulated based on the quantum circuit in 
(a) are presented in (g) for $r=0.6$, where 
	the concurrence between the system qubits and between the system qubit and ancilla are shown with dotted and continuous red lines 
respectively, the dot-dashed and continuous black lines present the 
linear entropy of the system qubit and of the ancilla, and the continuous blue 
line corresponds to the three-tangle entanglement.}
\label{fig:correlation}
\end{figure}
\end{widetext}

The experimental implementation on the IBM quantum processor is carried out using three qubits, as
shown in Fig.~\ref{fig:correlation}(a). As before, we average over 8192 realizations.
We perform the complete quantum state tomography of the two-qubit system in the post-selected subspace at various different values of time $t$. This is done using a set of seven experiments on the system qubits $\mathrm{q}$ and $\mathrm{q}'$, followed by $\sigma_z$-measurements of all three qubits 
as shown in Fig.~\ref{fig:correlation}(a) -- see Supplementary Note \ref{supp-note-5}  for further details. At the end of each of these tomography measurements, the populations
are obtained as $p_{kl}=P_{0kl}/\sum_{m,n=0}^1 P_{0mn}$, where $k,l \in\{0,1\}$, from which the desired density operator of the system qubits $\rho_{\rm q,q'}^{(0)}$ in the post-selected subspace is obtained. To study the entanglement dynamics we use the concurrence~\cite{Wootters-prl-1998,Horodecki-rmp-2009} as a measure of entanglement, given by
$\mathcal{C}_{\rm q,q'}^{(0)}= {\rm max} \{ 0, \sqrt{\lambda_1}-\sqrt{\lambda_2}-\sqrt{\lambda_3}-\sqrt{\lambda_4}\}$, where $\lambda_i$'s are the eigenvalues of the operator $\rho_{\rm q,q'}^{(0)} (\sigma_y \otimes \sigma_y) (\rho_{\rm q,q'}^{(0)})^{\star} (\sigma_y \otimes \sigma_y)$ written in decreasing order.

The change of concurrence with time is then observed for different values of $r$, as shown in Fig.~\ref{fig:correlation}(b)-(d).
For $r=0$ we have checked that the dynamics is unitary and there is no variation in the entanglement values. In this case, the standard result that the entanglement is not changed by local operations is confirmed.
However, for 
$0<r<1$, the concurrence is found to be oscillating, which is clearly seen in Fig.~\ref{fig:correlation}(b), while for
$r>1$ the Hamiltonian $H_{\mathrm q}$ governing local evolution ceases to obey the $\mathcal{PT}$ symmetry,
and the entanglement gradually decays with time, as shown in Fig.~\ref{fig:correlation}(d). For $r=1$
we find the same theoretical asymptotic scaling as for distinguishability $\mathcal{C}^{(0)}_{\rm q, q'} \sim t^{-\delta}$, where $\delta =2$. To compare with the experiment, again we restrict the time to $t \in [1,3]$, and find 
 $\delta = -1.71 \pm 0.01$ from the theoretical curve and $\delta = -1.93 \pm 0.27$ from the measured data (see inset).
In Fig.~\ref{fig:correlation}(f) we present the corresponding theoretical curves for the time-variation of concurrence for a wider range of $r$ parameters. 

The obtained variation in concurrence under a local operation contradicts at first sight the well-known property of monotonicity of entanglement. To make this effect even more striking, we have performed another experiment where we observe the increase in entanglement between the qubits $\mathrm{q}$ and $\mathrm{q}'$ under the 
	action of the $\mathcal{PT}$-symmetric non-Hermitian 
Hamiltonian. Specifically, at $t=0$ we prepare the state $\cos (\vartheta) |\Phi^{-}\rangle_{\rm q, q'} - i \sin (\vartheta) |\Psi^{+}\rangle_{\rm q, q'}$, where $|\Psi^{\pm}\rangle_{\rm q, q'} = (|01\rangle_{\rm q, q'} \pm |10\rangle_{\rm q, q'} )/\sqrt{2}$ and $|\Phi^{\pm} \rangle_{\rm q, q'}= (|00\rangle_{\rm q, q'} \pm |11\rangle_{\rm q, q'} )/\sqrt{2}$ are the standard maximally entangled Bell states.  The angle $\vartheta$ defines the concurrence $|\cos (2 \vartheta)|$
of this state. For the experiment - shown in Fig. \ref{fig:correlation} (e) - we took $\vartheta=59.185^0$, yielding a concurrence of 0.475 at $t=0$. The preparation of this state is done by single- and two- qubit gates acting on the two qubits; the ancilla is not involved and remains separate in the state $|0\rangle_{\rm a}$. Next, we simulate the action of the non-Hermitian Hamiltonian for $0 \geq t <2$ and $r=0.3$, $r=1$, $r=1.3$, see Fig. \ref{fig:correlation}(e). In the first case, the entanglement increases up to approximately 0.8 (and would continue to oscillate at longer times), while for $r=1$, $r=1.3$ it decreases monotonously.

\subsection*{Entanglement correlations between system and ancilla}

The simulation of non-Hermiticity by the dilation method allows us to get an in-depth understanding of this phenomenon. Let us look at the complete picture in the eight-dimensional Hilbert space of this tri-partite system
(initialized in the state $|0\rangle \otimes |\Phi^{+}\rangle$), where, 
as we have seen in Fig.~\ref{fig:correlation}(a), one of the system qubits $\rm q$ along with the ancillary qubit `$\rm a$' undergo the unitary evolution 
$U_{\mathrm a,q}$. The relevant correlations for the ensuing analysis are plotted in Fig.\ref{fig:correlation}(g) for $r=0.6$.
We define the single-qubit reduced states by tracing out the other qubits $\rho_{i}={\rm Tr}_{j,h}[\rho_{i,j,h}]$, while the two-qubit reduced density operators
are obtained by a single partial trace operation $\rho_{i,j}={\rm Tr}_h [\rho_{i,j,h}]$,
where the three qubits are labeled by $i,j,h \in \{\rm a,q,q'\}$. The concurrence associated with the state $\rho_{i,j}$ is denoted by $\mathcal{C}_{i,j}$ and it is calculated using the formula for mixed two-qubit states mentioned earlier. It is interesting to note that ${\rm q}$ and ${\rm q'}$ are always in the permutation symmetric subspace of the 
two-qubit Hilbert space as one of the qubits evolves under $H_{\mathrm q}$ (see Supplementary Note \ref{supp-note-3}). Therefore, it is enough to observe any one of the system qubits. Analyzing first the single-qubit states, we find that the 
single qubit reduced density operators $\rho_{\rm q}$ and $\rho_{\rm q'}$ 
remain maximally mixed all through the evolution, with a constant value of linear entropy $s_{\mathrm{q}}=1-\rm{Tr}(\rho_{q}^2)=0.5$.

Next, we observe that the concurrences $\mathcal{C}_{\rm a,q}$ and $\mathcal{C}_{\rm a,q'}$
 between the ancilla and the respective system qubits, {\it i.e.} $\rm a$ and $\rm q$ (or $\rm a$ and $\rm q'$) always remain zero.
This shows that the dynamics under $\mathbb{U}_{\mathrm a,q,q'}$ does not develop bipartite correlations between the 
respective system qubits and the ancilla qubit. 
Therefore, the  creation of a tripartite
correlation between the system and the ancilla can happen only through entangling correlations between the 
two-qubit reduced state 
of $\rm q, \rm q'$ and the ancilla. 
To quantify this tripartite correlation we use the three-tangle for pure  states~\cite{Horodecki-rmp-2009}
$\tau_{\rm{a,q,q'}}=\mathcal{C}_{\rm{a:q,q'}}^2 -\mathcal{C}_{\rm{a:q}}^2-\mathcal{C}_{\rm{a:q'}}^2$, 
or
	\begin{equation}
	\tau_{\rm{a,q,q'}}=\mathcal{C}_{\rm{q:a,q'}}^2 -\mathcal{C}_{\rm{q:a}}^2-\mathcal{C}_{\rm{q:q'}}^2, \label{eq:tangle1}
	\end{equation}
	where in the last equation we used the invariance of the tangle under permutations. As shown by Eq.~(\ref{eq:tangle1}), 
	the maximum value of the three-tangle is 
	obtained in the absence of concurrence between the individual components. 
	Here $\mathcal{C}_{\rm{q:a}}\equiv \mathcal{C}_{\rm{q,a}}$ and $\mathcal{C}_{\rm{q:q'}}\equiv\mathcal{C}_{\rm{q,q'}}$ are the
	concurrences of the two-party reduced states $\rho_{\rm a,q}$ and
	 $\rho_{\rm q,q'}$. 
	 The first term on the right hand side of Eq.~(\ref{eq:tangle1}) is 
	 the square of concurrence between the 
	 bi-partitions $\rho_{\rm q}:\rho_{\rm a,q'}$, where one partition 
	 consists of the qubit  ${\rm q}$ while the other partition
	 is formed by the ancilla ${\rm a}$ and the system qubit ${\rm q'}$.
	For a pure three-qubit state $\rho_{\rm a,q,q'}$, the quantity $\mathcal{C}_{\rm{q:a,q'}}$ is effectively related to the mixedness 
	of its bipartitions. More specifically, the square of concurrence 
	between the 
	partitions $\rho_{\rm q}$ and $\rho_{\rm a,q'}$ is twice the linear 
	entropy of the reduced density operator of either partition,
	given by $2(1-\rm{Tr}\rho_q^2)$ or $2(1-\rm{Tr}\rho_{\rm a,q'}^2)$.
	 We now know from the simulated dynamics that the linear entropy  
		$s_{\rm q(q')}=1-\rm{Tr}\rho_{\rm q(q')}^2=0.5$, therefore at all 
		times $\mathcal{C}_{\rm{q:a,q'}}^2=1$. 
	Further, as shown in Fig. \ref{fig:correlation}(g), there is no bipartite entanglement between the ancilla and the respective system qubits q (or q'), which implies that $\mathcal{C}_{\rm q,a(a,q')}=0$.
	From Eq.~(\ref{eq:tangle1}), we obtain,
	\begin{eqnarray} \label{eq:tangle2}
	\tau_{\rm a,q,q'} &=& 1- \mathcal{C}_{\rm q,q'}^2. 
	\end{eqnarray}
	Thus, the three-tangle among system qubits and
	ancilla and the concurrence between the system qubits are complementary to each other (see Supplementary Note \ref{supp-note-3}). By permuting the partitions in Eq.~(\ref{eq:tangle1}), it is easy to obtain $\tau_{\rm a,q,q'} = 2s_{\rm a}$, where $s_{\mathrm{a}}=1-\rm{Tr}(\rho_{\mathrm{a}}^2)$. 
    The unitary $\mathbb{U}_{\rm a,q,q'}$, which induces a local non-Hermitian drive of qubit ${\rm q}$
    in the post-selected subspace of the ancilla, is in fact a non-local operation on the system qubit ${\rm q}$ and the ancilla ${\rm a}$.
    Under $\mathbb{U}_{\rm a,q,q'}$, as the ancilla entangles and dis-entangles
with the joint state of the system qubits, we see the resulting
 oscillations of various correlations in time. These oscillating correlations
with $r$-dependent characteristic times, when postselected in the ancilla subspace, produce an apparent violation of entanglement monotonicity.
	 While we observe experimentally the variation in entanglement
	under local operations in only one post-selected subspace, other subspaces of the same system also witnesses 
	similar patterns for the variation of entanglement under local operations as shown in Supplementary Figure \ref{fig:ent_subspace1}.

\section*{Conclusion \label{sec:discussion}}
We have realized a quantum simulation of a single-qubit under a non-Hermitian Hamiltonian, observing the $\mathcal{PT}$-symmetry breaking as the exceptional point is crossed and the associated change in distinguishability.
The use of a quantum processor for the simulation has the advantage that more complex scenarios can be studied, such as 
a bipartite system with one of the qubits driven by a non-Hermitian Hamiltonian. In this case we observe the violation of the entanglement monotonicity no-go result from standard quantum mechanics. We also note that, while our method relies on dilation by the use of an ancilla, another approach to non-Hermitian evolution exists, where the dimension of the Hilbert space remains the same but the standard inner product is modified (see Methods). These two methods can be put in an exact correspondence - the metric used in the latter approach can be identified as the operator $\eta^2 + \mathbb{I}$ from the dilation approach.

The simulation of phenomena governed by $\mathcal{PT}$-symmetry at the single-quantum level open up several novel perspectives. Our scheme provides a systematic way of studying  more complex non-Hermitian many-qubit systems. It is important to realize that for a system of $N$ qubits the overhead in the width of the circuit is just one ancilla qubit.
	For example, it would be straightforward to generalize to the study of entanglement that we have performed to one non-Hermitian qubit and $N-1$ Hermitian ones,  in which case the depth of the circuit remains equal to 8. Furthermore, because we have access to the quantum regime, our scheme enables the study of quantum fluctuations. Since these systems are open -  connected to a source of energy providing gain and reservoir for dumping this energy - they naturally will lead to new insights into quantum thermodynamics.


\section*{Methods  \label{sec:methods}}
\subsection*{Simulating the non-Hermitian Hamiltonian in the dilated space \label{sm:hqa}}

The single-qubit evolution under a general time-dependent non-Hermitian Hamiltonian $H_{\mathrm{q}}(t)$ is obtained in a certain subspace of an ancilla-qubit system  
undergoing a unitary evolution generated by $\mathcal{H}_{\mathrm{a,q}}(t)$. The Hamiltonian $\mathcal{H}_{\mathrm{a,q}}(t)$ in a four-dimensional Hilbert 
space can be obtained from $H_{\mathrm{q}}$ by Naimark dilation~\cite{yang-science-2019}. 
Using this method, we can write the Hamiltonian $\mathcal{H}_{\mathrm{a,q}}(t)$ in the form: 
\begin{equation} \label{sm:eq:hqa}
 \mathcal{H}_{\mathrm{a,q}}(t)=  \mathbb{I} \otimes \Lambda(t)   + \sigma_y \otimes  \Gamma(t), 
\end{equation}
with
\begin{eqnarray} 
 \Lambda(t) &=& \left[H_{\mathrm{q}}(t) + i\frac{d\eta(t)}{dt} \eta(t) + \eta(t) H_{\mathrm{q}}(t) \eta(t) \right] \mathrm{M}^{-1}(t),  \label{sm:eq:hqa1}\\
  \Gamma(t) &=& i\left[ H_{\mathrm{q}}(t) \eta(t) - \eta(t) H_{\mathrm{q}}(t) - i\frac{d\eta(t)}{dt}  \right] \mathrm{M}^{-1}(t), \label{sm:eq:hqa2}\\
  \eta(t) &=& (\mathrm{M}(t)-\mathbb{I})^\frac{1}{2}, \qquad \qquad \rm{and} \label{sm:eq:hqa3} \\
  \mathrm{M}(t) &=& \mathrm{T} \exp \left[-i \int_{0}^{t}d \tau H_q^{\dag}(\tau )\right]\mathrm{M}(0) \widetilde{\mathrm{T}} \exp\left[i \int_{0}^{t}d\tau H_q (\tau)\right], \nonumber \\ \label{sm:eq:hqa4}
\end{eqnarray}
 where $\eta(t)$ and $\textrm{M}(t)$ are Hermitian operators; $\mathrm{T}$ and $\widetilde{\mathrm{T}}$ are time-ordering and anti-time-ordering operators respectively,
 and $\mathbb{I}$ is the $2 \times 2$ identity operator. The Hamiltonian  $\mathcal{H}_{\mathrm{a,q}}(t)$ can be obtained as follows. First, the equations Eq. (\ref{sm:eq:hqa3},\ref{sm:eq:hqa4}) for $\eta (t)$ and $M(t)$ reflect the invariance of the norm of $\ket{\Psi}_{\mathrm{a,q}}(t)$ in the form Eq. (\ref{bigpsi}) on the dilated space under evolution (see also the discussion about metric below). Then, the Schr\"odinger equations $i (d/d t) \ket{\Psi}_{\mathrm{a,q}} =  \mathcal{H}_{\mathrm{a,q}}(t) \ket{\Psi}_{\mathrm{a,q}}$,  and $i(d/dt)\ket{\psi(t)}_{\mathrm q}= H_{\mathrm q}(t)\ket{\psi(t)}_{\mathrm q}$ together with Eq. (\ref{sm:eq:hqa}) and Eq. (\ref{bigpsi}) produce a linear system of equations in the unknown operators $\Lambda (t)$ and $\Gamma (t)$, 
 \begin{eqnarray}
 \Lambda (t) - i \Gamma (t) \eta (t) &=& H_{\rm q}(t) , \\
 \Lambda (t) \eta (t) + i\Gamma (t) &=& i \frac{d \eta(t)}{dt} + \eta (t) H_{\rm q}(t) .
 \end{eqnarray}
 By multiplying the second equation to the right with $\eta (t)$ and with $-\eta^{-1}(t)$ and adding the results to the first equation we obtain immediately the solution as given by Eqs. (\ref{sm:eq:hqa1}, \ref{sm:eq:hqa2}). Note also that for Hermitian Hamiltonians $H_{\rm q}$ the second term in Eq. (\ref{sm:eq:hqa}), which is the qubit-ancilla interaction, becomes zero.

\subsection*{Initial conditions}

 To obtain an explicit form of $\mathcal{H}_{\mathrm{a,q}}(t)$, one should choose the operator $\textrm{M}(t)$ at time $t=0$ such that $\textrm{M}(t)-\mathbb{I}$ is 
 positive for all $t$ in the desired time interval. As a preliminary choice, we can take
 \begin{equation}
  \mathrm{M}(t=0)=\mathrm{M}_0=m_0 \times \mathbb{I}, \label{sm:eq:m01}
 \end{equation}
 where $m_0 > 1$, may be chosen arbitrarily. Further, we obtain the eigenvalues of $\textrm{M}(t)$ in the desired time interval. Fixing the value of $r$,
 at any arbitrary time $t$, the eigenvalues of $\textrm{M}(t)$ are labelled as $\mu_1(t)$ and $\mu_2(t)$, 
  from where we numerically obtain $\mu_{\mathrm{min}}(t)=\mathrm{min}\{ \mu_1(t), \mu_2(t) \}$. Interestingly, for $r=0$, $H_{\textrm{q}}$ 
  is Hermitian and
 \[ \mathrm{M}(t)=m_0 \times \mathbb{I} \quad \quad \forall \, t, \]
 with eigenvalues $m_0$, which is the maximum value that $\mu_{min}$ can assume.  Therefore, for any arbitrary $r$ and $t$, $m_0/\mu_{\mathrm{min}} \geq 1$.
 Thus, at $t=0$, $\textrm{M}(t)$ is chosen to be, 
 \begin{equation}
  \mathrm{M}(t=0)=\mathrm{M}_0= \frac{m_0}{\mu_{\rm min}}f \times \mathbb{I}, \label{sm:eq:m02}
 \end{equation}
where $f>1$, which ensures that $\textrm{M}(t)-\mathbb{I}$ remains positive for all $t$. From Eq.~\ref{sm:eq:hqa3} we have
 \begin{equation}
  \eta(t=0)= \left( \frac{m_0}{\mu_{\rm min}}f -1 \right)^{\frac{1}{2}} \times \mathbb{I} = \eta_0 \times \mathbb{I}. \label{sm:eq:m03}
 \end{equation}
The dynamics of the total ancilla-qubit system under
$\mathcal{H}_{\mathrm{a,q}}(t)$ is obtained from the Schr\"odinger equation
\begin{equation}
i\frac{d}{dt}\ket{\Psi(t)}_{\mathrm{a,q}}= \mathcal{H}_{\mathrm{a,q}}(t)\ket{\Psi(t)}_{\mathrm{a,q}},
\end{equation}
whose solution is given by
\begin{equation}\label{bigpsi_methods}
\ket{\Psi(t)}_{\mathrm{a,q}} = \ket{0}_{\mathrm a}\ket{\psi(t)}_{\mathrm q} + \ket{1}_{\mathrm a}\ket{\tilde{\psi}(t)}_{\mathrm q},
\end{equation}
where $\ket{\psi(t)}_{\mathrm q}$ is the solution of $i\frac{d}{dt}\ket{\psi(t)}_{\mathrm q}= H_{\mathrm q}\ket{\psi(t)}_{\mathrm q}$, and 
$\ket{\tilde{\psi}(t)}_{\mathrm q}=\eta(t)\ket{\psi(t)}_{\mathrm q}$. At $t=0$ the state of the total system is
\begin{eqnarray}\label{sm:bigpsi2}
\ket{\Psi(0)}_{\mathrm{a,q}} &=& \ket{0}_{\mathrm a}\ket{\psi(0)}_{\mathrm q} + \ket{1}_{\mathrm a} \ket{\tilde{\psi}(0)}_{\mathrm q}, \nonumber \\
                            &=& \ket{0}_{\mathrm a}\ket{\psi(0)}_{\mathrm q} + \ket{1}_{\mathrm a} \eta(0) \ket{\psi(0)}_{\mathrm q},  \nonumber \\
                            &=& \left( \ket{0}_{\mathrm a} +  \eta_0 \ket{1}_{\mathrm a} \right) \otimes \ket{\psi(0)}_{\mathrm q} 
                            = \ket{\psi(0)}_{\mathrm a} \otimes \ket{\psi(0)}_{\mathrm q},  \nonumber \\ \label{eq:initial}
\end{eqnarray}
which is a separable state of the ancilla $\ket{\psi(0)}_{\mathrm a}$ and the system qubit $\ket{\psi(0)}_{\mathrm q}$.
For preparing the initial state Eq. (\ref{eq:initial}) the ancilla is taken in one of the eigenvectors of $\sigma_z$, say $\vert 0 \rangle_{\rm a}$. This is then subjected to a rotation by an angle $\theta$ around the $y$ axis,  $R_y(\theta) = \exp (-i\theta \sigma_y/2)$, where, $\theta=2 \tan^{-1} \eta_0$. This leads to
\begin{eqnarray}
 \vert \psi(0) \rangle_{\rm a} &=& \cos \frac{\theta}{2} \vert 0 \rangle_{\rm a} + \sin\frac{\theta}{2} \vert 1 \rangle_{\rm a} 
  = \cos \frac{\theta}{2} \left( \vert 0 \rangle_{\rm a} + \tan\frac{\theta}{2} \vert 1 \rangle_{\rm a} \right),  \nonumber \\
    &=&  \frac{1}{\sqrt{\eta_0^2+1}} \left( \vert 0 \rangle_{\rm a} + \eta_0 \vert 1 \rangle_{\rm a} \right),
\end{eqnarray}
which is the initial state as defined by the protocol. On the other hand, the qubit $\mathrm{q}$ may 
be initialized in any arbitrary state $\ket{\psi(0)}_{\mathrm q}$. For the case of a single qubit, as discussed in the first part of the paper, we considered two different 
values of the state of the qubit: (i) $\ket{\psi(0)}_{\mathrm q} = \vert 0 \rangle$ and (ii) $\ket{\psi(0)}_{\mathrm q} = \vert 1 \rangle$. The same formalism applies also in the case of the two-qubit system discussed in the second part of the paper, in which case the qubit-qubit system is initialized in a maximally entangled Bell state $\ket{\psi(0)}_{\mathrm q} \rightarrow \ket{\Phi^{+}} = \frac{1}{\sqrt{2}}(\vert 00 \rangle + \vert 11 \rangle )$.
  
\par 
For instance, in case i) we choose $m_0=2$, and $f=1.01$. 
At $r = 0.6$ in the time range $t \in [0,8]$, we obtain $\eta_0 = 1.7436$ and $\theta = 2.1001$ (radians).
At the exceptional point, i.e. $r = 1$, in the same time range $t \in [0,8]$, we get $\eta_0 = 16.1112$ and $\theta = 3.0176$ radians.
Further, for $r = 1.3$, $\mu_{\rm min}$ is obtained separately for various time intervals to increase the probability of success. This then led to 
different values of $\eta_0$ and hence $\theta$ for each time point.

\subsection*{The metric \label{sm:metric}}

Non-Hermitian quantum dynamics can be alternatively formulated by using Hilbert spaces with a modified bra vector, resulting in a redefinition of the inner product \cite{Bender2007,Moussa2016,Nori2019}. Here we make an explicit connection with this approach, showing that the Hermitian operator $M(t)=\eta(t)^2 + \mathbb{I}$ can be identified as the metric that plays a key role in this formalism. 

Indeed, from Eq. (\ref{bigpsi_methods}) we can calculate the norm of the dilated vector $\ket{\Psi(t)}_{\mathrm{a,q}}$,
\begin{equation}
_{\mathrm{a,q}}\braket{\Psi(t)}{\Psi(t)}_{\mathrm{a,q}} = ~_{\mathrm{q}}\braket{\psi(t)|M(t)}{\psi(t)}_{\mathrm{q}} \label{eq:fullnorm}
\end{equation}
This norm has to be conserved during the time evolution. By taking the time-derivative of Eq. (\ref{eq:fullnorm}) and using Eq. (\ref{eq:extendedSchr}) we get
\begin{equation}
i \frac{d}{dt} M(t) = H_{q}^{\dag}(t) M(t) - M(t) H_{q}(t).
\end{equation}
This is the defining relation for the metric \cite{Nori2019}. Note that this can also be obtained in a straigthforward way from Eq. (\ref{sm:eq:hqa4}). Thus, in this approach to non-Hermitian quantum mechanics for every vector $|\psi(t)\rangle_{\textrm{q}}$ in the Hilbert space we define the covector as $_{\textrm{q}}\langle \psi(t)|M(t)$, which ensures that the inner product $_{\mathrm{q}}\braket{\psi(t)|M(t)}{\psi(t)}_{\mathrm{q}}$ from Eq. (\ref{eq:fullnorm})
has the meaning of a conserved probability.

For the particular case of the Hamiltonian $H_{q}$ studied in this work, the metric $M(t)$ can be obtained analytically by employing the properties of   $2 \times  2$ matrices (see also Supplementary Equation (\ref{Supplementary-eq-3})). 
In the $\mathcal{PT}$-symmetric phase we obtain an exact formula for the metric,
\begin{eqnarray}
M(t)/M_{0} &=& \frac{1}{1-r^{2}} \left[1 - r^{2} \cos\left( 2\sqrt{1-r^2}t \right)
\right]\mathbb{I}  \nonumber \\
& & + \frac{r}{\sqrt{1-r^2}} \sin \left(2\sqrt{1-r^2}t\right) \sigma_{z} \nonumber \\
& & + \frac{r}{1-r^2}\left[1 - \cos 2\sqrt{1-r^2}t\right] \sigma_{y}
\label{eq:M}
\end{eqnarray}
One can check also that $M(t)$ is positively defined for $r<1$, while for $r=0$ we recover the standard Hermitian quantum mechanics with $M(t)=M_{0}$. The result above Eq. (\ref{eq:M}) can be also obtained from the generic formula for the metric, as per
\cite{Nori2019}, for parameters  $A= 0$, $B= -r/(1-r^2 )$, $C= 1/(1-r^2)$, and $D=0$.

It is interesting to remark how the main problem of non-Hermitian quantum mechanics, that of non-conservation of probability, has been dealt with in completely different ways: either by the introduction of a metric and modifying the inner product, or, in the dilation method, by adding an ancilla that absorbs the excess population.

\subsection*{Evolution and measurement in the dilated space \label{main:operator}}

Let us consider the evolution of an arbitrary state of a two-qubit system under the 
Hamiltonian $\mathcal{H}_{\mathrm{a,q}}(t)$ in Eq.~(\ref{sm:eq:hqa}),
\[ i\frac{d}{dt}\ket{\psi(t)}_{\mathrm{a,q}}= \mathcal{H}_{\mathrm{a,q}}(t)\ket{\psi(0)}_{\mathrm{a,q}}, \]
where $\psi(0)$ is the initial state at $t=0$. This may also be written as 
\begin{eqnarray}
 \ket{\psi(t)}_{\mathrm{a,q}} &=& \rm{T} \exp \left[ i \int_0^{{t}} \mathcal{H}_{\mathrm{a,q}}(\tau) d\tau \right]\ket{\psi(0)}_{\mathrm{a,q}} \nonumber  \\ 
 &=& U_{\mathrm{a,q}}(t) \ket{\psi(0)}_{\mathrm{a,q}}, 
 \end{eqnarray}
where $\rm{T}$ is the time ordering operator.
For a given set of values of $r$ and $t$, it is useful to obtain an explicit form of the unitary operator $U_{\mathrm{a,q}}(t)$. 
This is done by observing the respective $\mathcal{H}_{\mathrm{a,q}}(t)$ evolutions of the complete set of two-qubit basis states.
To find $U_{\mathrm{a,q}}(t)$ we solve the Schr\"odinger equation numerically for different initial states,
\begin{equation}
	\begin{cases}
		i\frac{d}{dt}\ket{\psi_{kl}(t)}_{\mathrm{a,q}}= \mathcal{H}_{\mathrm{a,q}}(t)\ket{\psi_{kl}(t)}_{\mathrm{a,q}} \\
		\ket{\psi_{kl}(0)} = \ket{kl}_{\mathrm{a,q}} & k,l = 0,1 .
	\end{cases}
\end{equation}
where $\ket{00}_{\mathrm{a,q}}$, $\ket{01}_{\mathrm{a,q}}$, $\ket{10}_{\mathrm{a,q}}$, and $\ket{11}_{\mathrm{a,q}}$ correspond to the 
complete set of basis vectors in the four-dimensional Hilbert space. 
The system qubit and the ancilla are initialized in all four bases states respectively and then
evolved numerically under $\mathcal{H}_{\mathrm{a,q}}(t)$ for a given time. 
Then, after solving this equation for $\ket{\psi_{00}(t)}$, $\ket{\psi_{01}(t)}$, $\ket{\psi_{10}(t)}$, and $\ket{\psi_{11}(t)}$ we 
obtain the closed form of the unitary operator at an arbitrary time $t$, given by
\begin{eqnarray}
U_{\mathrm{a,q}}(t) &=& \ket{\psi_{00}(t)}_{\mathrm{a,q}}\bra{00} + \ket{\psi_{01}(t)}_{\mathrm{a,q}}\bra{01} \nonumber \\
& & + \ket{\psi_{10}(t)}_{\mathrm{a,q}}\bra{10} + \ket{\psi_{11}(t)}_{\mathrm{a,q}}\bra{11}.
\end{eqnarray}
For various different values of time, $U_{\mathrm{a,q}}(t)$ is obtained, which is a general unitary operator 
in the four dimensional Hilbert space. Each $U_{\mathrm{a,q}}(t)$ at a given time is then decomposed numerically in the form of single-qubit rotations
and two-qubit controlled-NOT gates, as shown in Fig.(1e).

This quantum circuit decomposition gives rise to $U_{\mathrm{num}}(t)$, whose operation is very close to the 
theoretical $U_{\mathrm{a,q}}(t)$. To characterize this, we calculate the error function  
$\mathrm{err}_{U}(t) = ||U_{\mathrm{a,q}}(t) - U^{\rm num}(t)||_2/||U_{\mathrm{a,q}}(t)||_2$, with the 2-norm defined by $||A||_2=\sqrt{\lambda_{\rm max}}$, where $\lambda_{\rm max}$ is the
largest eigenvalue of the matrix $A^{*}A$. 
Here 
$U^{\rm num}(t)$ is an unitary operator generated by the circuit in the inset of Fig.~\ref{fig:pop}(a), where the parameters $\alpha,\beta,\gamma$ of $\mathcal{U}_{\rm{q(a)}}^j(\alpha,\beta,\gamma)$ are chosen to minimize the expression $||U_{\mathrm{a,q}}(t) - U^{\rm num}(t)||_2$. 
Typically, we find
$\mathrm{err}_{U}(t) = ||U_{\mathrm{a,q}}(t) - U^{\rm num}(t)||_2/||U_{\mathrm{a,q}}(t)||_2$ to be of the order of $10^{-4}$, which 
demonstrates the high accuracy of our $U_{\mathrm{a,q}}$ implementation.
The accuracy with which our gate decomposition and the 
$U_{\mathrm{a,q}}(t)$ operator match with each other is presented by an example data set in the Supplementary Table 1. $U_{\mathrm{a,q}}(t)$ for arbitrary values of ($r,t$)
 and the corresponding $U^{\mathrm{num}}(t)$ can be obtained from a GitHub code repository~\cite{github}.

\subsection*{Quantum state reconstruction \label{methods:section2}}

For single-qubit tomography we take 4-1=3 measurements, corresponding to the set of Pauli operators $\sigma_x, \sigma_y, \sigma_z$. 
For higher dimensional quantum systems of $n$ qubits we 
need $2^{2n}-1$ measurements corresponding to combinations of $\sigma_x, \sigma_y, \sigma_z$ and the identity matrix $\mathbb{I}$ of the two qubits.
Therefore, a complete quantum state tomography of a two-qubit system requires a set of (16-1) experiments, which correspond to determining the expectation values of all the two-qubit operators formed by products of Pauli operators and the identity. 
In the present work, we need only to examine the post-selected subspace of the total system with the ancilla in state $\vert 0 \rangle$. Therefore we 
circumvent the complexities of three-qubit tomography by restricting our measurement to a $4\times4$ block of the complete $8\times8$ 
three-qubit density operator.

We perform a complete quantum state tomography of the system qubits by applying the following seven operators, namely
$ \mathbb{T}_1=\mathbb{I} \otimes \mathbb{I},\: \mathbb{T}_2=H \otimes \mathbb{I}, \:  \mathbb{T}_3=R_x(\pi/2) \otimes \mathbb{I}, \:  \mathbb{T}_4=\mathbb{I} \otimes H, \:  \mathbb{T}_5=\mathbb{I} \otimes R_x(\pi/2), \:  \mathbb{T}_6=(\mathbb{I} \otimes H) \mathrm{CNOT}, \:  \mathbb{T}_7=(\mathbb{I} \otimes R_x(\pi/2)) \mathrm{CNOT} $. The application of each of these operators is followed by the measurement in the $\sigma_z$ bases and post-selection of the desired subspace. 
Thus, in each of these experiments, we measure all three qubits, and obtain eight diagonal elements $p_{\rm i,j,k}=\vert c_{\rm i,j,k} \vert^2$. Finally, the corresponding populations of the two-qubit reduced density operator in the postselected subspace with ancilla in state $|0\rangle_{\rm a}$ are given by
\begin{equation}
 p_{\rm j,k}^{(0)}=\frac{p_{0,\rm{j,k}}}{\sum_{\rm{j,k=0}}^1p_{0,\rm{j,k}}}.
\end{equation}
Next, these populations are corrected for measurement errors, and the post-selected two-qubit density operators obtained further undergo convex optimization~\cite{cvx, gb08} (see Supplementary Note \ref{supp-note-5} for further details). 
\section*{Data availability}
The data that support the findings of this study are
available from authors upon reasonable request. 

\section*{Code availability} 
The codes used for the simulations can be found in the GitHub repository \cite{github}.

\section*{Acknowledgements}
We acknowledge support from the Foundational Questions Institute Fund (FQXi) via the Grant No. FQXi-IAF19-06, from the EU Horizon 2020 research and innovation  programme  (grant  agreement  no. 862644, FET  Open  QUARTET),  and from the Academy of Finland through project no. 328193 and through the “Finnish
Center of Excellence in Quantum Technology QTF” project no. 312296. In addition, we would like to thank the Scientific  Advisory  Board  for  Defence  (Finland)  and Saab.  
We acknowledge the use of IBM Quantum services for this work. The views expressed are those of the authors, and do not reflect the official policy or position of IBM or the IBM Quantum team. 

\section*{Author contributions}
SD and GSP initiated the project and obtained the key analytical results. AM designed the quantum circuits and performed the simulations on the IBM Q machine with input from SD.  All authors discussed the results. SD and GSP wrote the manuscript, with contributions also from AM.
\section*{Corresponding authors}
Correspondence to Shruti Dogra or Gheorghe Sorin Paraoanu.
\section*{Competing interests}
The authors declare no competing interests.
%

\newpage
\begin{widetext}
\begin{center} 
\Large{\textbf{Supplementary Material}}
 \end{center}

\section{Supplementary Note 1: The non-Hermitian Hamiltonian \label{sm:hamiltonian}}

\subsection{$\mathcal{PT}$-symmetry}
Consider the generic qubit non-Hermitian Hamiltonian (natural units $\hbar =1$)
\begin{equation}
H_{\mathrm q} = \sigma_{x} + ir \sigma_{z}, 
\label{eq:hamiltonian_generic}
\end{equation}
where $r$ is a real parameter and $\sigma_{x}$ and $\sigma_{z}$ are the Pauli matrices. 
Since this is essentially a spin-1/2 system, the parity operator 
is $\mathcal{P} = \sigma_{x}$ and the time-inversion operator is the complex conjugation operator $\mathcal{T} = \star$. 
$\mathcal{PT}$-symmetry requires that $H_q$ and $\mathcal{PT}$ should share a common set of eigenvectors, which may be easily 
verified by operating $\mathcal{PT}$ on the eigenvectors of $H_q$. We have three cases: 

{\it Case I: $0 \leq r < 1$.}
In this situation $\sqrt{1-r^2}$ is real and we obtain the eigenvalues of $H_{\mathrm q}$ as $\pm \sqrt{1-r^2}$, with the corresponding eigenvectors
\[ \vert \psi_{\pm} \rangle = \frac{1}{\sqrt{2}}\begin{pmatrix} ir \pm \sqrt{1-r^2} \\ 1 \end{pmatrix}. \]
Acting with the $\mathcal{PT}$ operator we
get 
\[ \mathcal{PT} \vert \psi_{\pm} \rangle = (-ir\pm\sqrt{1-r^2})\vert \psi_{\pm} \rangle . \]
Thus, $\vert \psi_{\pm} \rangle$ is also an eigenvector of $\mathcal{PT}$, which 
implies that,  in the given range of the parameter $r$, the system Hamiltonian $H_q$ is $\mathcal{PT}$-symmetric. In particular, at $r=0$ the Hamiltonian reduces to $\sigma_{x}$ and the eigenvectors are real -  the time-reversal in the $\mathcal{PT}$ operator does not play any role, and the only remaining action is that of $\sigma_x$ (identical to the Hamiltonian). 


{\it Case II: $r > 1$.}
In this situation $\sqrt{1-r^2}$ is purely imaginary; to put in evidence this we can write it as $i\sqrt{r^2-1}$. The eigenvectors corresponding  to the eigenvalues $\pm i\sqrt{r^2-1}$ are 
\[ \vert \psi_{\pm} \rangle = \frac{1}{\sqrt{2}}\begin{pmatrix} ir \pm i \sqrt{r^2-1} \\ 1 \end{pmatrix}, \]
and acting with the time-reversal operator will result now in a change of sign of the square root term.
Specifically,
\[ \mathcal{PT} \vert \psi_{\pm} \rangle = -i (r\pm\sqrt{r^2-1})\vert \psi_{\mp} \rangle .\] 
In this case the $\mathcal{PT}$-symmetry is broken, and the action of the $\mathcal{PT}$ operator interchanges the eigenvalues. 

{\it Case III: $r = 1$.}
In this case the two eigenvalues coalesce into a single one, zero, with corresponding eigenvector
\[ \vert \psi_{0} \rangle = \frac{1}{\sqrt{2}}\begin{pmatrix} i \\ 1 \end{pmatrix}. \]
We get
\[ \mathcal{PT} \vert \psi_{0} \rangle = -i\vert \psi_{0} \rangle . \]
This is the exceptional point (EP).

\subsection{Bloch sphere representation of $\mathcal{PT}$-symmetry breaking}

A simple geometric representation of $\mathcal{PT}$-symmetry breaking can be obtained by using the Bloch sphere, see Supplementary Fig. \ref{fig:Bloch}. The eigenvalues derived above are represented by red and blue dashed lines, for $r \in [0,2]$. At $r=0$ the system Hamiltonian $H_q$ becomes $\sigma_x$ and its eigenvectors are orthogonal, 
lying along the $\pm x$-axes on the Bloch sphere (shown with blue and red colored dots). As $r$ increases from $0$ to $1$, the  
eigenvectors move closer to each other in the 
$xy$-plane and they are no longer orthogonal. Eventually
these dashed lines join together at the exceptional point $r=1$, corresponding 
to the coalescence of the eigenvectors of $H_s$. This merging of the eigenvectors 
is shown in Supplementary Fig. \ref{fig:Bloch}
as the green colored dot along the $y$-axis, with coordinates $(0,1,0)$. Going further to $r>1$, the pair of eigenvectors separates again, moving in the $yz$-plane of the Bloch sphere. For very large values of $r$, the diagonal terms of $H_s$ begin to dominate and the corresponding 
eigenvectors approach the eigenvectors of $\sigma_z$. Thus, the eigenvalues approach the $\pm z$-axes 
asymptotically, which is shown by the black dots at $(0,0,\pm1)$.

\begin{figure}[h]
 \centering
 \includegraphics[scale=1,keepaspectratio=true]{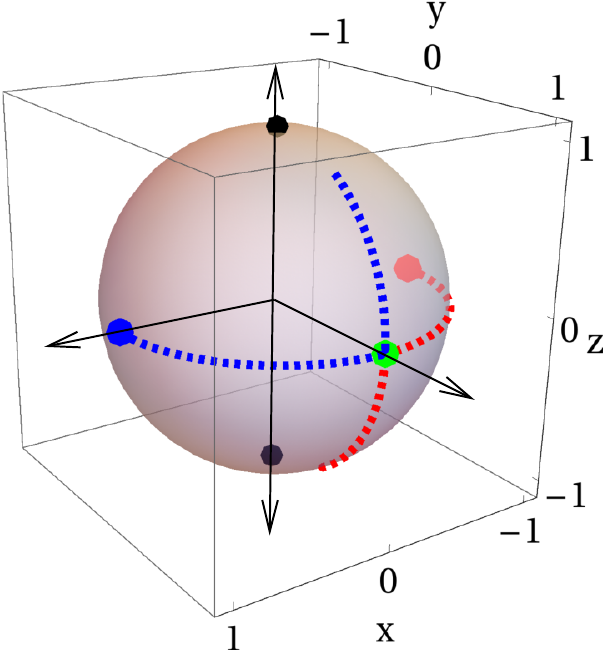}
 \caption{Bloch sphere representation of the eigenvectors of $H_q$, for various different values of $r \in [0,2]$.}
 \label{fig:Bloch}
\end{figure}

\subsection{Single qubit evolution under $H_{\rm q}$ \label{dynamics-hq}}

In Fig. 1 b) we have presented the evolution of the ground-state population under the action of
\begin{equation}
H_{\mathrm q} = \sigma_{x} + ir \sigma_{z},
\label{eq:hamiltonian_generic2}
\end{equation}
for various values of $r$. This can be obtained numerically as 
	\[ p_{0}(r,t)=|\langle 0 | e^{-i H_q t} |0\rangle|^2/(|\langle 0 | e^{-i H_q t} |0\rangle|^2+|\langle 1 | e^{-i H_q t} |0\rangle|^2), \] 
where we note that due to the non-Hermiticity the preservation of the norm is not guaranteed. 
This evolution can also be solved analytically. Indeed, we have 
\begin{equation}
e^{-iH_{\rm q}t} = \cos (\sqrt{1-r^2} t) \mathbb{I} - \frac{i}{\sqrt{1-r^2}} \sin (\sqrt{1-r^2}t) H_{\rm q}. \label{Supplementary-eq-3}
\end{equation}
On the single-qubit basis the action is
\[ |0\rangle  \longrightarrow \alpha_{0}(t) |0\rangle + \beta_{0}(t) |1\rangle, \quad \textrm{and} \quad |1\rangle \longrightarrow \alpha_{1}(t) |0\rangle + \beta_{1}(t) |1\rangle, \]
where
\begin{eqnarray}
\alpha_{0}(t) &=& \cos(\sqrt{1-r^2}t) + \frac{r}{\sqrt{1-r^2}} \sin(\sqrt{1-r^2}t) \nonumber \\
\beta_{0}(t) &=& \alpha_{1}(t) = -\frac{i}{\sqrt{1-r^2}} \sin(\sqrt{1-r^2}t) \nonumber \\
\beta_{1}(t) &=& \cos(\sqrt{1-r^2}t) - \frac{r}{\sqrt{1-r^2}} \sin(\sqrt{1-r^2} t). \label{SM:eq:evolution}
\end{eqnarray}
If the initial state is $|0\rangle$, this leads to the following expressions for the populations of the $\{|0\rangle, |1\rangle \}$ basis,
\begin{equation}
p_0(t)= \frac{|\alpha_{0}(t)|^2}{|\alpha_{0}(t)|^2+|\beta_{0}(t)|^2}  \quad \textrm{and} \quad  p_1(t)= \frac{|\beta_{0}(t)|^2}{|\alpha_{0}(t)|^2+|\beta_{0}(t)|^2}. \label{sm:eq:pop}
\end{equation}


\vspace{10mm}
\section{Supplementary Note 2: Evolution in the dilated space \label{sm:operator}}
\begin{table}[h!]
    \centering
    \small
    \begin{tabular}{|c|c|c|c||c|c|c|c|}
    	\hline 
    	$t$ & $p_{0}(t)$ & $p_{0}^{\rm num}(t)$ & $\mathrm{err}_{U}(t)$ & $t$ & $p_{0}(t)$ & $p_{0}^{\rm num}(t)$ & $\mathrm{err}_{U}(t)$ \\ & & & & & & & \\
     	\hline 
    	 $0.5$  & $0.8613$  & $0.8613$  & $2.0\times 10^{-5}$  & $4.5$  & $0.8315$  & $0.8315$  & $9.0\times 10^{-5}$ \\ & & & & & & &  \\
    	 $1.0$  & $0.6547$  & $0.6547$  & $6.7\times 10^{-6}$  & $5.0$  & $0.6250$  & $0.6250$  & $1.0\times 10^{-5}$ \\ & & & & & & &  \\
    	 $1.5$  & $0.4535$  & $0.4535$  &  $1.7\times 10^{-5}$ & $5.5$  & $0.4242$  & $0.4242$  & $5.0\times 10^{-6}$ \\ & & & & & & &  \\
    	 $2.0$  & $0.2495$  & $0.2495$  & $2.0\times 10^{-5}$  & $6.0$  & $0.2191$  & $0.2191$  & $2.0\times 10^{-5}$ \\ & & & & & & &  \\
    	 $2.5$  & $0.0518$  & $0.0518$  & $8.0\times 10^{-5}$  & $6.5$  &  $0.0300$ & $0.0300$  & $2.0\times 10^{-5}$ \\ & & & & & & &  \\
    	 $3.0$  & $0.0695$  & $0.0695$  & $6.4\times 10^{-6}$  & $7.0$  &  $0.1278$ & $0.1278$  & $7.6\times 10^{-5}$ \\ & & & & & & &  \\
    	 $3.5$  & $0.7314$  & $0.7314$  & $1.7\times 10^{-5}$  & $7.5$  & $0.8220$  & $0.8221$  & $1.0\times 10^{-4}$ \\ & & & & & & &  \\
    	 $4.0$  & $0.9951$  & $0.9951$  & $3.7\times 10^{-4}$  & $8.0$  & $0.9821$  & $0.9822$  & $1.7\times 10^{-4}$ \\
    	\hline 	
    \end{tabular}
    \caption{Ground state population and errors at $r=0.6$.  We present the population of the ground state $p_{0}(t)$ calculated from the evolution under $H_{\rm q}$ and the corresponding numerical values $p_{0}^{\rm num}(t)$ obtained by using the quantum gate decomposition of $U_{\mathrm{a,q}}(t)$. We also show the error function $err_U$ which characterizes the mismatch between the two evolution operators.}
    \label{table1}
\end{table}

From the main text, the operator in the dilated space is given by,
\begin{eqnarray}
U_{\mathrm{a,q}}(t) &=& \ket{\psi_{00}(t)}_{\mathrm{a,q}}\bra{00} + \ket{\psi_{01}(t)}_{\mathrm{a,q}}\bra{01} 
 + \ket{\psi_{10}(t)}_{\mathrm{a,q}}\bra{10} + \ket{\psi_{11}(t)}_{\mathrm{a,q}}\bra{11}.
\end{eqnarray}
For various different values of time, $U_{\mathrm{a,q}}(t)$ is obtained, which is a general unitary operator 
in a four-dimensional Hilbert space. Each $U_{\mathrm{a,q}}(t)$ at a given time is then decomposed numerically in the form of single-qubit rotations
and two-qubit controlled-NOT gates as shown in Fig.(1e) of the maintext.
This quantum circuit decomposition gives rise to $U_{\mathrm{num}}(t)$, whose operation is very close to the 
theoretical $U_{\mathrm{a,q}}(t)$. The accuracy with which our gate decomposition and the 
$U_{\mathrm{a,q}}(t)$ operator match with each other is presented here by an example data set.
Supplementary Table~\ref{table1} shows a comparison between the  
theoretically expected values obtained from evolution with $H_{\rm q}$ and the numerically obtained values obtained by simulating the results using the quantum circuit decomposition of 
$U_{\mathrm{a,q}}(t)$. Here $p_{0}(t)$ and $p_{0}^{\rm num}(t)$ are the corresponding ground-state populations of the qubit at various different values of time $t$.
To further characterize this, we calculate the error function  
$\mathrm{err}_{U}(t) = ||U_{\mathrm{a,q}}(t) - U^{\rm num}(t)||_2/||U_{\mathrm{a,q}}(t)||_2$, with the 2-norm defined by $||A||_2=\sqrt{\lambda_{\rm max}}$, where $\lambda_{\rm max}$ is the
largest eigenvalue of the matrix $A^{*}A$. Here 
$U^{\rm num}(t)$ is the unitary operator generated by the circuit in the inset of Fig.1(a) of the main text, where parameters $\alpha,\beta,\gamma$ of $\mathcal{U}_{\rm{q(a)}}^j(\alpha,\beta,\gamma)$ are chosen to minimize the expression $||U_{\mathrm{a,q}}(t) - U^{\rm num}(t)||_2$. It is clearly apparent from the table that the theoretically expected and 
numerically obtained values of the ground state populations at various different times match exactly up to at least three decimal places, which 
shows the high accuracy level of our $U_{\mathrm{a,q}}$ implementation.
 
\section{Supplementary Note 3: Dynamics of entanglement \label{supp-note-3}}
In this section we focus on the dynamics of entanglement in the two experimental setups (single qubit and ancilla and two qubits and ancilla).

\subsection{One-qubit system}
We observed non-Hermitian $\mathcal{PT}$-symmetric ($0 < \rm{r} <1$) and non-$\mathcal{PT}$-symmetric (r $\geq 1$) dynamics of the system qubit as shown in the main text.
Considering the system qubit and the ancilla undergoing the unitary dynamics generated by $\mathcal{H}_{\mathrm{a,q}}(t)$ in the dilated space, non-local 
correlations between the system and the ancilla are developed in time. By performing quantum tomography we obtain the  concurrence between the system and the ancilla shown in Supplementary Fig.~\ref{fig:ent2q} with red dots. The theoretically expected dependence is shown with continuous black line. Interestingly, the  
concurrence oscillate in time for $0 < \rm{r} <1$, which may be interpreted as the information retrieval of the system.

\begin{figure}[ht]
\centering
\includegraphics[scale=1,keepaspectratio=true]{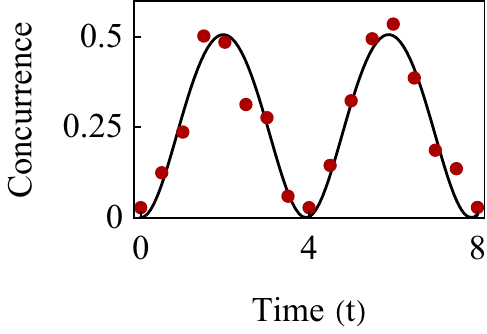}
\caption{Concurrence between system and ancilla qubit, as they both together undergo the dynamics 
generated by the dilated Hamiltonian $\mathcal{H}_{\rm a,q}(t)$ for r$=0.6$. The red dots are experimental data, while the black continuous line is the expected ideal theoretical result.}
\label{fig:ent2q}
\end{figure}

\subsection{Two-qubit system}

{\it Permutation symmetry of the two-qubit state}
\vspace{1mm}

We show here that the permutation symmetry (interchange of qubits $\rm{q} \leftrightarrow\rm{q'}$) of the initial Bell state $|\Phi^{+}\rangle=\frac{1}{\sqrt{2}}(|00\rangle + |11\rangle)$ is preserved under time evolution. This is not {\it a priori} obvious, since $H_{\rm q}$ acts locally only on qubit $\rm{q}$.
The analytical form of the two-qubit state at an arbitrary time $t$  is given by
 \begin{eqnarray}
  |\Phi(t)\rangle &=& (e^{-i H_{\rm q}t} \otimes \mathbb{I}) \; |\Phi^{+}\rangle \nonumber \\
  &=& \frac{1}{\sqrt{2}\mathcal{N}(t)}(\alpha_0(t)|00\rangle + \beta_0(t)|10\rangle +\alpha_1(t)|01\rangle +\beta_1(t)|11\rangle),
 \end{eqnarray}
where $\mathcal{N}(t)=\sqrt{\sum_{i=0}^1(|\alpha_i|^2+|\beta_i|^2)}$ is the normalization constant and 
coefficients $\alpha_i(t)$ and $\beta_i(t)$ are defined in Eqs.~(\ref{SM:eq:evolution}). Since $\alpha_1(t)=\beta_0(t)$, we have,
\begin{equation}
 |\Phi(t)\rangle =\frac{1}{\sqrt{2}}(\alpha_0(t)|00\rangle + \beta_0(t)(|10\rangle +|01\rangle) +\beta_1(t)|11\rangle).
\end{equation}
It is now easy to see that interchanging the qubits does not alter $|\Phi(t)\rangle$. 

\vspace{1mm}
{\it Entanglement restoration}
\vspace{1mm}

\begin{figure}[h!]
 \centering
 \includegraphics[scale=1,keepaspectratio=true]{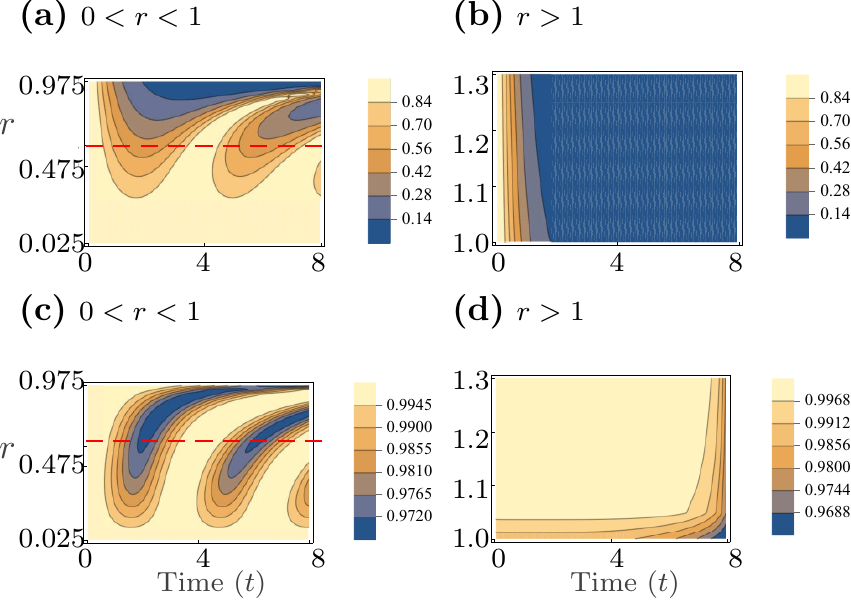}
 \caption{Concurrence between the system qubits q and q' as a function of r and t. Parts (a),(b) corrsepond to 
 the subspace postselected with ancilla in state $|0\rangle$, while (c),(d) correspond to the subspace with ancilla in state $|1\rangle$.}
 \label{fig:ent_subspace1}
\end{figure}

It has been shown in Fig. (3) of the main text that the concurrence between the system qubits 
varies with time when one of the system qubits undergoes a local evolution generated by a non-Hermitian Hamiltonian. 
This situation is experimentally realized in a dilated space of two system qubits and one ancilla, such that the 
dynamics of our interest lies in the post-selected subspace corresponding to ancilla in state $\vert 0 \rangle$. 
The results from the simulation in Supplementary Fig. \ref{fig:ent_subspace1}(a,b) depict the oscillations of the concurrence between 
the system qubits for $\rm{r} < 1$ and decay for $\rm{r}>1$ respectively in the desired post-selected subspace. 
A similar pattern of variation in concurrence is also observed in the post-selected subspace with ancilla in 
state $\vert 1 \rangle$, as shown in parts (c,d) of Supplementary Fig. \ref{fig:ent_subspace1}. However, in the complete eight 
dimensional space of system qubits and ancilla, these three-qubits undergo a unitary dynamics and entanglement 
does not vary under local operations.

While we observe the variation in entanglement
under local operations in one of the post-selected subspaces, other subspace of the same system also witnesses 
similar patterns for the variation of entanglement under local operations. As expected from standard Hermitian quantum  mechanics, in the combined picture of system and ancilla
the apparent violation of entanglement monotonicity no longer exists. A complete picture of correlations in this three-qubit system 
is presented in Supplementary Fig.~\ref{fig:ent3q}.
\begin{figure}
 \centering
 \includegraphics[scale=1,keepaspectratio=true]{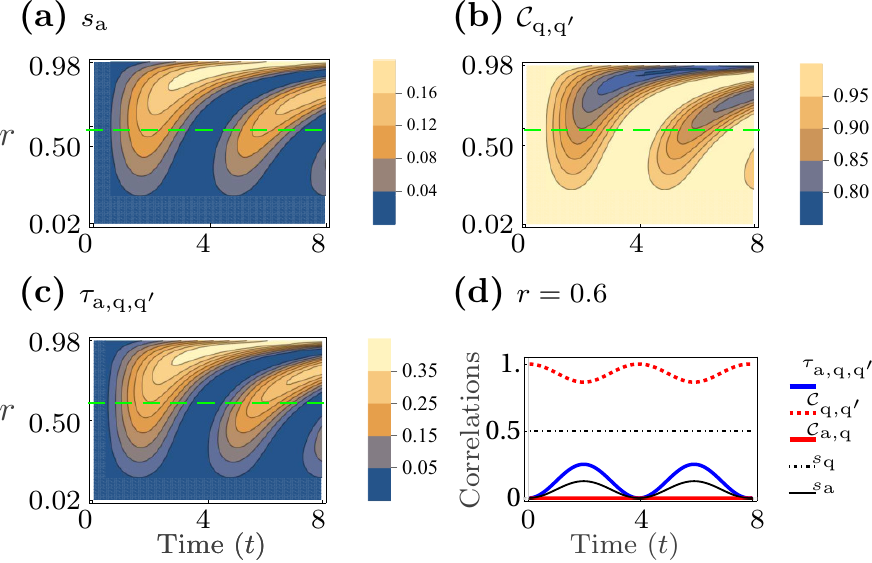}
 \caption{Dynamics (a) of the von Neumann entropy of the single-qubit reduced density matrix,
 (b) of the concurrence between $\mathrm{q}$ and $\mathrm{q}'$, and (c) of the three-tangle, shown over a range $r \in [ 0.02, 0.98]$.
 The plots in (d) show the dynamics of various quantities (specified at the right side of the figure) for $r=0.6$, which is marked by 
 a horizontal dashed green line in the previous contour plots (a), (b), (c).}
 \label{fig:ent3q}
\end{figure}
\section{Supplementary Note 4: Information retrieval and loss \label{supp-note-4}}

Recalling the single-qubit dynamics from Section~\ref{dynamics-hq}, the time-evolved state of 
the system $\rho(t)$ with the  non-Hermitian Hamiltonian $H_{\rm q}$ is given by~\cite{kohei-prl-2017}
\begin{equation}
 \rho(t) = \frac{e^{-i H_{\rm q}t} \rho(0) e^{i H_{\rm q}^{\dag}t}}{\rm{Tr} [e^{-i H_{\rm q}t} \rho(0) e^{i H_{\rm q}^{\dag}t}]},
\end{equation}
where $\rho(0)$ is the initial state of the system. Using the spectral decomposition, $\rho(0)=\sum_{mn} \rho_{mn} |\psi_m\rangle \langle \psi_n|$; 
in the eigenbasis of the Hamiltonian ($H_{\rm q}$) we obtain
\begin{equation}
 \rho(t) = \frac{\sum_{mn} \rho_{mn} e^{-i (E_m - E_n)t} |\psi_m\rangle \langle \psi_n|}{\sum_{mn} \rho_{mn} e^{-i (E_m - E_n)t} \langle \psi_n|\psi_m\rangle}. \label{eq:ReccTime_rhot}
\end{equation}
In the case of a single qubit system, $m,n \in \{+,-\}$ such that $E_{\pm}$ are the eigenvalues of $H_{\rm q}$ with corresponding eigenvectors $|\psi_{\pm}\rangle$
and the difference in the eigenvalues, $E_m - E_n= 2\sqrt{|1-r^2|}$. 
For $r \neq 0$, the eigenvectors $|\psi_{\pm}\rangle$ are not mutually orthogonal, which results in a time-dependent denominator (from now on, denoted by $N(t)$) 
in Supplementary Eq.~\ref{eq:ReccTime_rhot}. Supplementary Fig.~\ref{fig:norm} presents the variation of the normalization $1/N(t)$ with time $t$ for different 
values of $r$. 

If $r\in (0,1)$, both the numerator and denominator of Supplementary Eq.~\ref{eq:ReccTime_rhot} oscillate in time at 
the same rate. Thus there is information exchange between the system and the environment. The system retrieves the information from the environment in a time $T_R$, the so-called recurrence time. Therefore for any arbitrary $t$, $ N(t+T_R)= N(t)$, leading to
\begin{eqnarray}
 e^{-2i \sqrt{1-r^2} t} = e^{-2i \sqrt{1-r^2} (t+T_R)}, \nonumber
\end{eqnarray}
which gives
\begin{equation}
 T_R = \frac{\pi}{\sqrt{1-r^2}} \label{eq:ReccTime}.
\end{equation}
\begin{figure}
 \centering
 \includegraphics[scale=0.6,keepaspectratio=true]{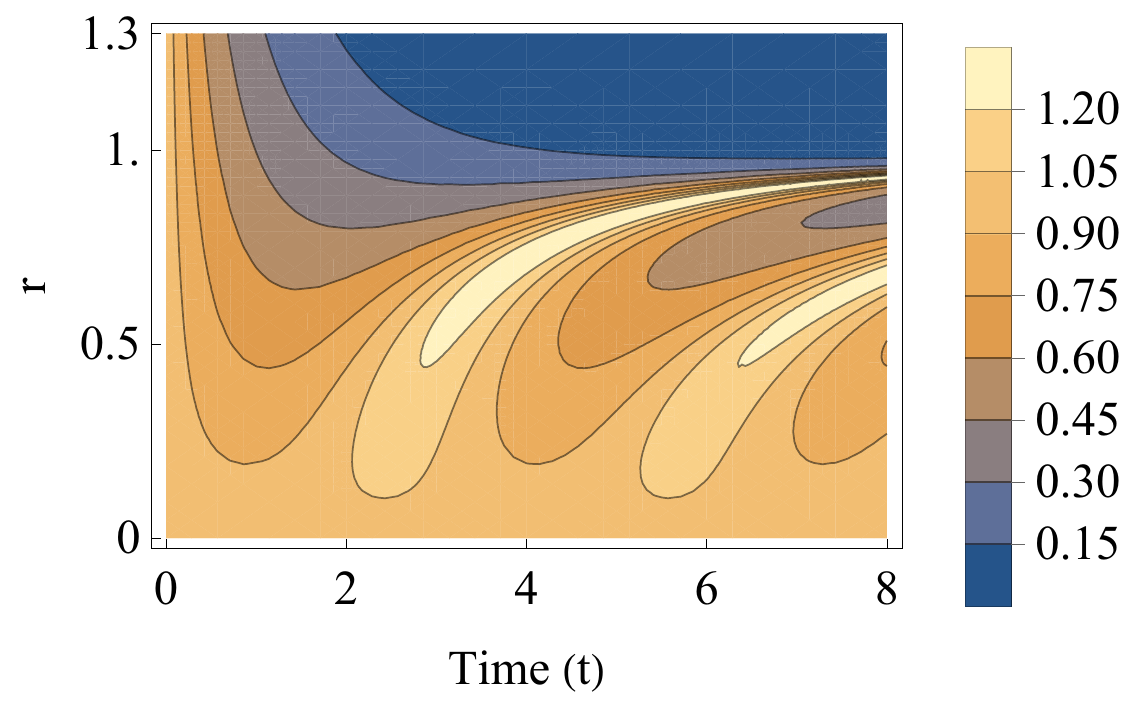}
 \caption{Variation of the norm $1/N(t)$ with time for different values of $r$, starting from the initial state
 	$\rho(0)=|0\rangle \langle 0|$. Oscillations may be seen 
 for $r\in(0,1)$ and a decaying pattern for $r>1$.}
 \label{fig:norm}
\end{figure}

For $r>1$, the Hamiltonian $H_{\rm q}$ does not exhibit $\mathcal{PT}$ symmetry. In this regime, 
the quantum system looses its information exponentially to the environment and never regains it back.  
The rate of loss of information may be quantified in terms of the decay time ($\tau_{\rm D}$) of 
the system. This may be interpreted as the time in which the norm $1/N(t)$ of the state $\rho(t)$ 
diminishes by $e$. Therefore we have
\begin{equation}
 \frac{1}{N(t+\tau_{\rm D})} = \frac{1}{eN(t)} \nonumber
\end{equation}
 or
 \begin{equation}
\tau_{\rm D} = \frac{1}{2\sqrt{r^2-1}}.  \label{eq:DecayTime}
\end{equation}

These properties of recurrence and decay for $r\in(0,1)$ and $r>1$ respectively may be observed 
in the measurement of various quantities, as shown in the main text.

The values of recurrence time ($T_{\rm R}$) and decay time ($\tau_{\rm D}$) are obtained experimentally by fitting the data using Supplementary Eq.~(\ref{sm:eq:pop}).
In the case $r<1$, $\sqrt{1-r^2}$ in Eqs.~(\ref{SM:eq:evolution}) is substituted by $\pi/T_{\rm R}$, while for $r>1$, $\sqrt{1-r^2}$ is replaced by $-i/2\tau_{\rm D}$.
These characteristic times are the properties of the Hamiltonian $H_{\rm q}$ as they emerge solely from the quantum state evolution. Interestingly, we encounter
these characteristic time scales in the study of various quantities. For instance, the measure of distinguishability, which is based on the trace distance 
between two arbitrary pairs of single-qubit states, oscillates at $T_{\rm R}$ for the case of $r < 1$ and decays at $\tau_{\rm D}$ for $r>1$. This
may be readily seen in the simulation in Fig. 2(a) of the main text, and also in the experimental curves for $r=0.6$ and $r=1.3$ in Fig. 2(c,e) of the main text. 
The same $T_{\rm R}$ and $\tau_{\rm D}$ emerge in the study of entanglement dynamics between two-qubits when one of these qubits is evolved under the non-Hermitian Hamiltonian $H_{\rm q}$. Indeed, these characteristic times appear in the experimental data from the main text, and they are well supported by the theoretical results of Fig. 3(f) in the main text and of Supplementary Fig.~\ref{fig:ent3q}.
An interesting scenario is presented in Supplementary  Fig.~\ref{fig:ent_subspace1}, where in the subspaces post-selected by ancilla in state $|0\rangle$ and $|1\rangle$ 
respectively, oscillations in the measure of concurrence for $0<r<1$ and decay for $r>1$ exhibit different patterns. The underlying cause is the evolution under different
non-Hermitian Hamiltonians, effective in the respective post-selected subspaces.
\\
In general, these oscillating and decaying patterns signify the  
information retrieval or information loss at $r$-dependent characteristic rates. These may be further interpreted to study the memory effects as in a 
non-Markovian dynamics and information decay in a Markovian dynamics respectively.

\section{Supplementary Note 5: Additional experimental information \label{supp-note-5}}

In this work we used the ibmq\_ourense computer from IBM Q Experience. In experiments with 2 qubits we employed the zeroth and the first qubits. In experiments with 3 qubits we used the zeroth, the first and the second qubits. In Supplementary Table \ref{tab:my_label} we list their characteristics. By cx:i-j we denote a CNOT gate with qubit i as control and j as target.

\begin{table}[h!]
\centering
\begin{tabular}{|c|c|c|c|c|c|}
\hline
Qubit  & T1($\mu$s)  & T2($\mu$s)  & Readout error(\%) & Single-qubit U2 error(\%)      & CNOT error(\%)\\ \hline
0      &  76         & 57          &  2.0        & 0.03                     & cx1-0: 0.62\\ \hline
1      &  75        & 28          &  2.7        & 0.05                     &                 \\ \hline
2      &  100         & 98          &  1.7        & 0.03                     & cx1-2: 0.84  \\ \hline
\end{tabular}
\caption{Parameters of the superconducting quantum processor used in this work.}
\label{tab:my_label}
\end{table}

Quantum state tomography of the system qubits is performed by applying seven operators, which fetch 
the matrix elements ($\rho^{exp}_{\rm row,col}$; $\rm{row, col} \in \{ 1,2,3,4\}$) of the experimental two qubit state $\rho^{exp}$,
as listed in Supplementary Table~\ref{table4}.
\begin{table}[h!]
	\centering
	\begin{tabular}{|c|c|c|}
		\hline & & \\
	$\mathbb{T}_1$      &  $\mathbb{I} \otimes \mathbb{I}$     &   $\rho^{exp}_{1,1},\rho^{exp}_{2,2},\rho^{exp}_{3,3},\rho^{exp}_{4,4}$  \\ \hline & & \\
	$\mathbb{T}_2$      &  $H \otimes \mathbb{I}$     &   $\rm{Re}(\rho^{exp}_{1,3}), \rm{Re}(\rho^{exp}_{2,4})$  \\ \hline & & \\
	$\mathbb{T}_3$      &  $R_x(\pi/2) \otimes \mathbb{I}$     &   $\rm{Im}(\rho^{exp}_{1,3}), \rm{Im}(\rho^{exp}_{2,4})$  \\ \hline & & \\
	$\mathbb{T}_4$      &  $\mathbb{I} \otimes H$     &   $\rm{Re}(\rho^{exp}_{1,2}), \rm{Re}(\rho^{exp}_{3,4})$  \\ \hline & & \\
	$\mathbb{T}_5$      &  $\mathbb{I} \otimes R_x(\pi/2)$     &   $\rm{Im}(\rho^{exp}_{1,2}), \rm{Im}(\rho^{exp}_{3,4})$  \\ \hline & & \\
	$\mathbb{T}_6$      &  $(H \otimes \mathbb{I}) \, \mathrm{CNOT}$     &   $\rm{Re}(\rho^{exp}_{1,4}), \rm{Re}(\rho^{exp}_{2,3})$  \\ \hline & & \\
	$\mathbb{T}_7$      &  $(R_x(\pi/2) \otimes \mathbb{I}) \, \mathrm{CNOT}$     &   $\rm{Im}(\rho^{exp}_{1,4}), \rm{Im}(\rho^{exp}_{2,3})$  \\ \hline
	\end{tabular}
\caption{Tomography operations and corresponding resulting elements of the two-qubit density operator.}
	\label{table4}
\end{table}

In this scheme, the first operator
($\mathbb{T}_1$) provides the populations ($4-1=3$ variables). The operators $\mathbb{T}_2$ and $\mathbb{T}_3$ project the real and imaginary parts of the single-quantum coherences of the second qubit (4 variables) on the diagonal positions 
of the density operator, which is then read by $\sigma_z$ measurements. In a similar way,  $\mathbb{T}_4$ and $\mathbb{T}_5$ provide the single-quantum coherences of the first qubit (4 variables).
In order to obtain the terms corresponding to the zero quantum coherence ($\rho_{2,3}^{exp}$) and double-quantum coherence ($\rho_{1,4}^{exp}$)  of the density operator, we use two-party correlations, which is a faster way to obtain these terms. Thus, the remaining 
four variables are obtained via operations $\mathbb{T}_6$ and $\mathbb{T}_7$.

\subsection{Experimentally obtained state of the system}
Multiple implementations (here 8192) of these seven tomography operations are performed on the system qubits (q and q'),
while at the end of the sequence, all three qubits: ancilla (a) and system qubits are measured. These measurements actually
provide the relative occupancy of the eight bases states of a three-qubit system at the end of each measurement,
which we call here `populations'. There is a possiblity of measurement errors associated with one or more of these populations,
such as, a single-qubit state, which is in state $|0\rangle$ might be wrongly measured as $|1\rangle$ or vice versa. 
For the three-qubit case, a correction matrix~\cite{steffen-science-2006, lucero-prl-2008} can be associated with each qubit,
 \begin{displaymath}
  F_i= \left( \begin{array}{ll}
        f_0^{(i)} & 1-f_1^{(i)} \\ 1-f_0^{(i)} & f_1^{(i)}
       \end{array} \right).
 \end{displaymath}
 where $f_{0(1)}^{(i)}$ is the probability by which a state $|0\rangle (|1\rangle)$ of the $i^{th}$ qubit is correctly identified as 
 $|0\rangle (|1\rangle)$, while $1-f_{0(1)}^{(i)}$ is the probability by which a state that is actually $|0\rangle (|1\rangle)$ is
 being wrongly considered as $|1\rangle (|0\rangle)$. Here $f_{0}^{(i)}$ and $f_1^{(i)}$ are the  measurement fidelities
of the bases states $|0\rangle$ and $|1\rangle$ respectively of the $i^{th}$ qubit ($i \in \{ a, q, q'\}$). Thus the correction matrix 
for a three-qubit system is given by,
\[ F= F_{\rm a} \otimes F_{\rm q} \otimes F_{\rm q'}. \]

In order to compensate the effect of errors induced by the measurements, the corrected populations ($p_{corr}$) are obtained by, 
$p_{corr}=F^{-1} p_m$, where $p_m$ stand for the experimentally measured populations of the eight bases
 states. Measurement fidelities of each of these qubits are measured independent of each other by measuring their respective bases states ($|0\rangle$ and $|1\rangle$).
We obtain the measurement fidelities:  $f_0^{(a)}=0.99$, $f_1^{(a)}=0.89$, $f_0^{(q)}=0.99$, 
$f_1^{(q)}=0.98$, $f_0^{(q')}=0.99$, $f_1^{(q')}=0.98$.
However, we need only to examine the post-selected subspace of the total system with the ancilla in state $\vert 0 \rangle$. Therefore we 
circumvent the complexities of three-qubit tomography by restricting our measurement to a $4\times4$ block of the complete $8\times8$ 
three-qubit density operator. This is elaborated in the next sub-section.
Using these intrinsic population values ($p_{corr}$), post-selected two-qubit density operators are obtained, 
which further undergo the convex optimization.

\subsection{Postselection}
Let us denote by $\rho_{\rm a, q, q'}$ the complete three-qubit density operator of the ancilla and the system qubits $\rho_{\rm a, q, q'}$,
\begin{equation}
 \rho_{\rm a, q, q'} = \sum_{\rm{i,j,k,l,m,n}=0}^{1} c_{\rm{i,j,k}} c_{\rm{l,m,n}}^{*} (|\rm{i}\rangle_{\rm a} |\rm{j} \rangle_{\rm q} |\rm{k}\rangle_{\rm q'}) (\langle \rm{l} |_{\rm a} \langle \rm{m} |_{\rm q} \langle \rm{n} |_{\rm q'}), \label{eq:rhofull}
\end{equation}
where $|\rm{i,j,k}\rangle_{\rm a,q,q'}$ (or $|\rm{l,m,n}\rangle_{\rm a,q,q'}$) form the three-qubit computational basis, and
$c_{\rm{i,j,k}}$ (or $c_{\rm{l,m,n}}$) are the corresponding coefficients of our three-qubit state, subject to the respective basis vectors.
A two-qubit reduced density operator of (say) system qubits is obtained by the partial trace operation, which eliminates the information about the 
ancilla qubit. The two-qubit reduced density operator is thus given by
\begin{equation}
    \begin{split}
       \rho_{\rm q, q'} & = \rm{Tr}_{\rm a} (\rho_{\rm a, q, q'}) \\
        & = \sum_{\rm{j,k,m,n}=0}^{1} \left[ c_{0,\rm{j,k}} c_{0,\rm{m,n}}^{*} ( |\rm{j} \rangle_{\rm q} |\rm{k}\rangle_{\rm q'}) (\langle \rm{m} |_{\rm q} \langle \rm{n} |_{\rm q'}) \right. \\
 \quad  & + \left. c_{1,\rm{j,k}} c_{1,\rm{m,n}}^{*} ( |\rm{j} \rangle_{\rm q} |\rm{k}\rangle_{\rm q'}) (\langle \rm{m} |_{\rm q} \langle \rm{n} |_{\rm q'}) \right]. \label{eq:rho2q}
    \end{split}
\end{equation}
Most often, we post-select the subspace with ancilla qubit in state $|0\rangle_{\rm a}$, and correspondingly, the two-qubit post-selected reduced density operator is given by,
\begin{eqnarray}
 \rho_{\rm q, q'}^{(0)} &=& \frac{1}{\mathcal{N}_{\rm qq'}^{(0)}} \sum_{\rm{j,k,m,n}=0}^{1} c_{0,\rm{j,k}} c_{0,\rm{m,n}}^{*} ( |\rm{j} \rangle_{\rm q} |\rm{k}\rangle_{\rm q'}) (\langle \rm{m} |_{\rm q} \langle \rm{n} |_{\rm q'}), \nonumber \\ \label{eq:rho2q0}
\end{eqnarray}
where $\mathcal{N}_{\rm qq'}^{(0)}$ is responsible for the normalization of this post-selected subspace.
To be more explicit, the two-qubit reduced subspace of our interest lies in a $4 \times 4$ block of the three-qubit state ($\rho_{\rm a,q,q'}$), which is shown colored in Supplementary  Fig.~\ref{fig:rhoaqq}. 
Interestingly, $\rho_{\rm q, q'}=\rho_{\rm q, q'}^{(0)}+\rho_{\rm q, q'}^{(1)}$. 
Further, the single-qubit reduced density operators in respective subspaces is obtained by partial trace of 
one of the qubits. For instance,
\[ \rho_{\rm q}^{(0)} = \rm{Tr}_{\rm q'} (\rho_{\rm q, q'}^{(0)}) \quad \rm{and} \quad \rho_{\rm q} = \rm{Tr}_{\rm q'} (\rho_{\rm q, q'}) .\]

As we described in the main text, a set of seven experiments are used to completely determine $\rho_{\rm q,q'}^{(0)}$.
At the end of each of these operators ($\mathbb{I}\otimes \mathbb{T}_{\iota}$, $\iota \in \{ 1,7 \}$), 
elements at the diagonal positions are directly accessed via $\sigma_z$ measurements. Thus, in each of these experiments, we measure all three qubits, and obtain eight diagonal elements $p_{\rm i,j,k}=\vert c_{\rm i,j,k} \vert^2$. Finally, the corresponding populations of the two-qubit reduced density operator in the postselected subspace, with ancilla in state $|0\rangle_{\rm a}$ is given by
\begin{equation}
 p_{\rm j,k}^{(0)}=\frac{p_{0,\rm{j,k}}}{\sum_{\rm{j,k=0}}^1p_{0,\rm{j,k}}}.
\end{equation}

\begin{figure}[]
 \centering
 \includegraphics[scale=0.7,keepaspectratio=true]{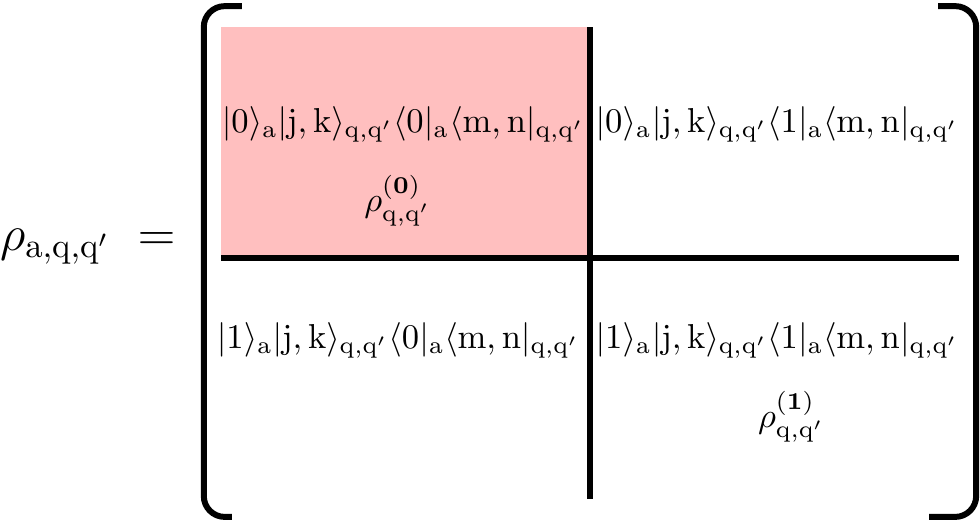}
 \caption{Abstract respresentation of the three-qubit density operator, with the desired post-selected subspace, 
 which is a $4 \times 4$ block highlighted with pink color.}\label{fig:rhoaqq}
\end{figure}

\end{widetext}
\end{document}